\documentclass[10pt,conference]{IEEEtran}

\usepackage[keeplastbox]{flushend}
\usepackage{subcaption}
\usepackage[rgb]{xcolor}
\usepackage{textcomp}
\usepackage{comment}
\usepackage{cite}
\usepackage{authblk}
\usepackage{amsmath}
\usepackage{resizegather}

\usepackage{epsfig}
\usepackage{endnotes}
\usepackage{graphicx}
\usepackage{float}
\usepackage{color} 
\usepackage{paralist}
\usepackage{listings}
\usepackage{html}

\usepackage{pgfplots}
\pgfplotsset{compat=newest}
\usetikzlibrary{decorations.pathmorphing}

\definecolor{bblue}{HTML}{4F81BD}
\definecolor{rred}{HTML}{C0504D}
\definecolor{ggreen}{HTML}{9BBB59}
\definecolor{ppurple}{HTML}{9F4C7C}

\usepackage{url}
\usepackage{breakurl}

\usepackage[flushleft]{threeparttable}

\usepackage{booktabs} 
\usepackage{clrscode} 
\usepackage{multirow}
\usepackage{algorithmic}
\usepackage[linesnumbered,ruled]{algorithm2e}
\usepackage{subcaption}
\usepackage{makecell}
\graphicspath{ {/} }
\newcommand{\ignore}[1]{}
\newcommand\textcode[1]{\texttt{#1}}

\newcommand{\Name}[1]{\textsf{Debreach}}
\newcommand{\SD}[1]{\textsf{SafeDeflate}}
\def\checkmark{\tikz\fill[scale=0.4](0,.35) -- (.25,0) -- (1,.7) -- (.25,.15) -- cycle;} 

\newcommand{\Ind}[1]{\hspace{#1ex}\hspace{#1ex}}

\definecolor{mygreen}{rgb}{0,0.6,0}
\definecolor{mygray}{rgb}{0.5,0.5,0.5}
\definecolor{mymauve}{rgb}{0.58,0,0.82}
\definecolor{navy}{RGB}{0,0,128}
\definecolor{cornflowerblue}{RGB}{100, 149, 237}
\definecolor{lightgoldenrod}{RGB}{238, 221, 130}
\definecolor{goldenrod}{RGB}{218, 165, 32}
\definecolor{tomato}{RGB}{255, 99, 71}
\definecolor{orangered}{RGB}{255, 69, 0}
\definecolor{lightcoral}{RGB}{240, 128, 128}
\definecolor{lightseagreen}{RGB}{32, 178, 170}
\definecolor{yellowgreen}{RGB}{154, 205, 50}
\definecolor{skyblue}{RGB}{135, 206, 250}
\definecolor{mistyrose}{RGB}{255, 228, 225}
\definecolor{bananamania}{rgb}{0.98, 0.91, 0.71}
\definecolor{blond}{rgb}{0.98, 0.94, 0.75}
\definecolor{champagne}{rgb}{0.97, 0.97, 0.81}
\definecolor{coralpink}{rgb}{0.97, 0.51, 0.47}
\definecolor{dkgreen}{rgb}{0,.6,0}
\definecolor{dkblue}{rgb}{0,0,.6}
\definecolor{dkyellow}{cmyk}{0,0,.8,.3}
\definecolor{pastelorange}{rgb}{1.0, 0.7, 0.28}

\usepackage{tikz}
\usetikzlibrary{positioning,arrows, shadows, calc, matrix, decorations.pathmorphing}
\usetikzlibrary{shadows}
\usetikzlibrary{shapes.geometric}
\usetikzlibrary{patterns}

\makeatletter 
\pgfdeclarepatternformonly[\LineSpace,\tikz@pattern@color]{my north east lines}{\pgfqpoint{-1pt}{-1pt}}{\pgfqpoint{\LineSpace}{\LineSpace}}{\pgfqpoint{\LineSpace}{\LineSpace}}%
{
    \pgfsetcolor{\tikz@pattern@color} 
    \pgfsetlinewidth{0.4pt}
    \pgfpathmoveto{\pgfqpoint{0pt}{0pt}}
    \pgfpathlineto{\pgfqpoint{\LineSpace + 0.1pt}{\LineSpace + 0.1pt}}
    \pgfusepath{stroke}
}
\makeatother 
\newdimen\LineSpace
\tikzset{
    line space/.code={\LineSpace=#1},
    line space=3pt
}

\makeatletter
\pgfdeclareshape{document}{
\inheritsavedanchors[from=rectangle] 
\inheritanchorborder[from=rectangle]
\inheritanchor[from=rectangle]{center}
\inheritanchor[from=rectangle]{north}
\inheritanchor[from=rectangle]{south}
\inheritanchor[from=rectangle]{west}
\inheritanchor[from=rectangle]{east}
\backgroundpath{
\southwest \pgf@xa=\pgf@x \pgf@ya=\pgf@y
\northeast \pgf@xb=\pgf@x \pgf@yb=\pgf@y
\pgf@xc=\pgf@xb \advance\pgf@xc by-7.5pt 
\pgf@yc=\pgf@yb \advance\pgf@yc by-7.5pt
\pgfpathmoveto{\pgfpoint{\pgf@xa}{\pgf@ya}}
\pgfpathlineto{\pgfpoint{\pgf@xa}{\pgf@yb}}
\pgfpathlineto{\pgfpoint{\pgf@xc}{\pgf@yb}}
\pgfpathlineto{\pgfpoint{\pgf@xb}{\pgf@yc}}
\pgfpathlineto{\pgfpoint{\pgf@xb}{\pgf@ya}}
\pgfpathclose
\pgfpathmoveto{\pgfpoint{\pgf@xc}{\pgf@yb}}
\pgfpathlineto{\pgfpoint{\pgf@xc}{\pgf@yc}}
\pgfpathlineto{\pgfpoint{\pgf@xb}{\pgf@yc}}
\pgfpathlineto{\pgfpoint{\pgf@xc}{\pgf@yc}}
}
}
\makeatother

\DeclareMathSymbol{"}{\mathalpha}{letters}{`"}

\begin{document}

\title{\Name{}: Mitigating Compression Side Channels via Static Analysis and Transformation}

\author[1]{Brandon Paulsen}
\author[1]{Chungha Sung}
\author[2]{Peter A.H. Peterson}
\author[1]{Chao Wang}
\affil[1]{University of Southern California, Los Angeles, CA, USA}
\affil[2]{University of Minnesota Duluth, Duluth, MN, USA}

\maketitle

\begin{abstract}
Compression is an emerging source of exploitable side-channel leakage
that threatens data security, particularly in web applications where
compression is indispensable for performance reasons.  Current
approaches to mitigating compression side channels have drawbacks in
that they either degrade compression ratio drastically or require too
much effort from developers to be widely adopted.
To bridge the gap, we develop \Name{}, a static analysis and program
transformation based approach to mitigating compression side
channels.  \Name{} consists of two steps.
First, it uses taint analysis to soundly identify flows of
sensitive data in the program and uses code instrumentation to
annotate data before feeding them to the compressor.
Second, it enhances the compressor to exploit the \emph{freedom to not
compress} of standard compression protocols, thus removing the
dependency between sensitive data and the size of the compressor's output.
Since \Name{} automatically instruments applications and does
not change the compression protocols, it has the advantage of being
non-disruptive and compatible with existing systems. 
We have evaluated \Name{} on a set of web server applications
consisting of 145K lines of PHP code. Our experiments show that, 
while ensuring leakage-freedom, \Name{} can achieve significantly higher
compression performance than state-of-the-art approaches.
\end{abstract}

\section{Introduction}

Compression is a technique for improving performance, especially in
web applications.  For example, the DEFLATE~\cite{rfc1951} compression
format in HTTP~\cite{rfc2616} is used by 70\% of the top one million
websites~\cite{compStats} because it reduces the size of web content
such as HTML, CSS and JavaScript by up to 70-80\%.  This not only
decreases latency and increases throughput, but also reduces energy
consumption~\cite{compenergy} for battery powered devices.
DEFLATE uses two techniques: Huffman coding~\cite{Huffman1952} and
LZ77 matching~\cite{lz77}, the latter of which is
particularly effective for web content.  It replaces repeated
strings with a \emph{reference} to an earlier copy in the input. For
example, if the input is \textit{``Bob is great, Bob is cool''}, the
output would be \textit{``Bob is great, $\langle 14,7 \rangle$cool''}
where the reference $\langle 14,7 \rangle$ is interpreted
as \textit{go back 14 bytes and then copy 7 bytes}. Web content has
many repeated strings, such as URLs and HTML tags; for example, the
string \textit{``wikipedia.org/wiki/''} appears 96 times on
Wikipedia's home page, which will be reduced in size by 85\%.

Unfortunately, dictionary compression in general, and LZ77 in
particular, introduces side channels that can be exploited
in \emph{partially chosen-plaintext attacks}~\cite{Kelsey2002,
Gluck:2, Thai2012, VanGoethem16Heist}.  In such a case, the attacker
feeds a \emph{guessed} text to the victim's application, which combines
the text with its own sensitive text before giving them to the
compressor. When the guessed text matches the sensitive text, the
compressed file will be smaller due to LZ77. Since encryption does not
hide size, this information can be leaked to an attacker even if the
compressed file is encrypted before transmission.  Consider the
sensitive text \emph{SSN: 123456789} and the guessed
texts \emph{SSN: 1234} and \emph{SSN: 1235}. The former will be a
complete match, but the latter will not include the last character
(5).  Since LZ77 references typically have the same size, the former will be one
byte smaller.

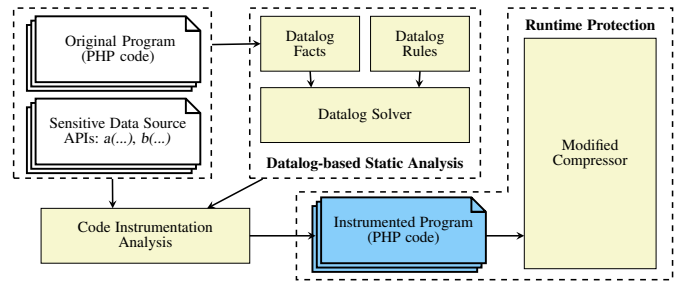
\begin{figure}
\vspace{1ex}
\centering
\scalebox{0.73}{\begin{tikzpicture}[font=\scriptsize] 
  \tikzstyle{arrow}=[thick,->,>=stealth,black]
  \tikzstyle{txt}=[above=5pt,right,text width=1.6cm]
  \tikzstyle{long_txt}=[above=5pt,right,text width=2cm]
  \tikzstyle{short_txt}=[above=5pt,right,text width=1.4cm]
    	
  
  \node[
  shape=document,
  double copy shadow={
    shadow xshift=-0.5ex,
    shadow yshift=-0.5ex
  },
  draw,
  fill=white,
  line width=1pt,
  minimum height=12mm,
  minimum width=3cm,
  align=center,
  inner sep=0,
  outer sep=0
  ] (I0) at (-1.0, 0.0) {{ \footnotesize Original Program} \\ \footnotesize (PHP code)};
  
  \node[
  shape=document,
  double copy shadow={
    shadow xshift=-0.5ex,
    shadow yshift=-0.5ex
  },
  draw,
  fill=white,
  line width=1pt,
  minimum height=12mm,
  minimum width=3cm,
  text width=30mm,
  align=center,
  inner sep=0,
  outer sep=0
  ] (I1) at (-1.0, -1.5) {{\footnotesize Sensitive Data Source} \\ \footnotesize APIs: \textsl{a(...)},  \textsl{b(...)}};

  \node[minimum width=36mm, minimum height=31mm, dashed, draw=black, thick, align=center] (D0) at (-1.1, -0.8) {};
  
  \node[draw=black, fill=champagne, minimum width=18mm, minimum height=10mm, text width=18mm, align=center, inner sep=0,outer sep=0] (DA0) at (2.5, 0.1) { \footnotesize Datalog \\ Facts};
  
  \node[draw=black, fill=champagne, minimum width=18mm, minimum height=10mm, text width=18mm, align=center, inner sep=0,outer sep=0] (DA1) at (4.5, 0.1) {\footnotesize Datalog \\ Rules};
  
  \node[draw=black, fill=champagne, minimum width=38mm, minimum height=10mm, text width=38mm, align=center, inner sep=0,outer sep=0] (DA2) at (3.5, -1.2) {\footnotesize Datalog Solver };
  
  \node[text width=4.0cm, align=center] at (3.5, -2.1) {\footnotesize \textbf{Datalog-based Static Analysis}};
 
  \node[minimum width=42mm, minimum height=31mm, dashed, draw=black, thick, align=center] (D1) at (3.5, -0.8) {};
  
  \node[draw=black, fill=champagne, minimum width=38mm, minimum height=10mm, text width=38mm, align=center, inner sep=0,outer sep=0] (INST) at (-0.5, -3.4) {\footnotesize Code Instrumentation\\ Analysis};
  
  \node[
  shape=document,
  double copy shadow={
    shadow xshift=-0.5ex,
    shadow yshift=-0.5ex
  },
  draw,
  fill=skyblue,
  line width=1pt,
  minimum height=12mm,
  minimum width=30mm,
  text width=30mm,
  align=center,
  inner sep=0,
  outer sep=0
  ] (R) at (4.2, -3.3) {{\footnotesize Instrumented Program}\\ \footnotesize(PHP code)};

   \node[draw=black, fill=champagne, minimum width=10mm, minimum height=42mm, text width=24mm, align=center, inner sep=0,outer sep=0] (COMP) at (7.6, -1.9) {\footnotesize Modified \\ Compressor};
 
   \draw[dashed, draw=black, thick] (6.1, 0.75) -- (9.1, 0.75) -- (9.1, -4.2) -- (2.25, -4.2) -- (2.25, -2.6) -- (6.1, -2.6) -- (6.1, 0.75);
   \node[text width=3.0cm, align=center] at (7.6, 0.45){\footnotesize \textbf{Runtime Protection}};
  
   \draw [arrow] (D0.26) -- (DA0.west); 
   \draw [arrow] (DA0.south) -- (DA2.153);
   \draw [arrow] (DA1.south) -- (DA2.27);
   \draw [arrow] (D1.220) -- (INST.24);
   \draw [arrow] (D0.south) -- (INST.140);
   \draw [arrow] (INST.east) -- (2.6, -3.4);
   \draw [arrow] (5.7, -3.4) -- (6.4, -3.4);

\end{tikzpicture}}
\caption{The overall flow of our \Name{} method.}
\label{fig:flow}
\vspace{-2ex}
\end{figure}

Compression side channels were first investigated by
Kelsey~\cite{Kelsey2002}, who found a potential attack against
encryption that later was adapted to 
the real world.
For example, Rizzo and Duong~\cite{Thai2012} proposed an attack named
CRIME for the widely-used TLS (successor of the now-deprecated SSL); in response to
the risk and due to lack of better solution, the community
accepted the solution of disabling TLS compression
entirely~\cite{compTLS1.3}, which is unfortunate.
Gluck et al.~\cite{Gluck:2} proposed BREACH, another refinement of
Kelsey's technique, which exploits HTTP compression; in response,
HTTP/2 altered the compression algorithm to prevent 
side-channel attacks on header data at the cost of
compressibility~\cite{HTTP2Comp}.
Both CRIME and BREACH require a passive eavesdropping point on the
encrypted connection to measure the encrypted data size; Vanhoef and
Goethem~\cite{VanGoethem16Heist, van2016}, on the other hand, lifted such a
requirement by estimating the size through the time taken for a
request to complete.

Despite the severity of such security threats, there does not yet exist a general
approach that provides sound guarantees about leakage-freedom
while maintaining acceptable levels of compression and
run-time performance.
To fill the gap, we propose \Name{}, an approach to
mitigating compression side channels in web server applications.
\Name{} can provide security guarantees about protecting
arbitrary web content as opposed to protecting, e.g., only
security tokens.
Furthermore, it does not require developers to manually identify flows
of sensitive data in the program; instead, it uses a 
static analysis to track the sensitive data flow, and based on the
analysis, transforms the server program to allow automated annotation
of sensitive data at run time.  In addition to automation, the \Name{} method
provides leakage-free guarantees and high compression performance
at the same time.

The overall flow is shown in Fig.~\ref{fig:flow}, where the input
consists of the program and a set of sensitive APIs. For example, if
an API function retrieves sensitive entries from a database, it will
be provided as a sensitive data source.
Inside \Name{}, a static taint analysis is used to identify flows of
sensitive data from sources to sinks, i.e., \emph{echo} statements
that construct the HTML file to be sent to the client. The analysis is
designed to be sound in that it guarantees to include all 
sensitive data flows.
The result is used to rewrite the program code such that it can, at
run time, insert annotations for sensitive data before feeding them to
the enhanced compressor.

Our taint analysis uses the general framework of
declarative program analysis~\cite{whaley2004, lam2005,
livshits2005, mangal2015}.  In this framework, the PHP program is
first traversed to produce a set of \emph{Datalog facts}, which encode
the control and data flow structures of the program as well as the
sensitive data sources.  The fixed-point computation required by taint
analysis is codified in a set of \emph{inference rules}, which are
combined with the facts to form the entire Datalog program.  We then solve
the Datalog program using our Python-based solver which is 
optimized for solving these rules, the output of which is a sound
overapproximation of all sensitive flows from sources to sinks.
Finally, we perform a code instrumentation analysis to determine the
optimal points to annotate the sensitive data, to ensure that
data are properly marked when they reach the compressor.
The instrumented program (PHP code), combined with our enhanced 
DEFLATE compressor, can skip LZ77 and instead use Huffman
coding only for the marked data.

In contrast, none of the previous works on mitigating compression side
channels can provide soundness guarantees about protecting arbitrary
data.
The mitigation adopted in HTTP/2~\cite{HTTP2Comp}, for example,
protects only header data but not the payload, while approaches based
on randomly masking sensitive data protect only security
tokens~\cite{django}, but not data embedded in JavaScript and
HTML~\cite{Karakostas16CTX}.
Size-randomizing techniques, while easily applicable, have also been shown
to be ineffective~\cite{Karakostas16}.  Finally, other approaches that
exploit the freedom to not compress~\cite{regexcomp, SafeDeflate}
either cannot protect arbitrary data~\cite{regexcomp} or do not
provide soundness guarantees~\cite{SafeDeflate}. In addition, none of
the existing techniques can perform safe compression and at the same
time avoid degrading the performance to unacceptable levels.
\Name{}, in contrast, is fully automated in generating sensitive data 
annotations and the accuracy is almost as high as annotations
created by experts manually.  

We have implemented \Name{} and evaluated it on a set of server
applications consisting of 145K lines of PHP code.  
The goal is to eliminate the dependence between compression
performance and sensitive data since it may be inferred or observed by attackers.
%
%
Our experiments show that \Name{} outperforms both \emph{SafeDeflate}, a
state-of-the-art technique that does not even provide leakage-free
guarantee, and Huffman coding, another technique that disables LZ77
entirely, in terms of compression performance.
Our experiments also show that \Name{} prevents side-channel leaks in
all applications, including several that \emph{SafeDeflate} cannot prevent.

To sum up, this paper makes the following contributions:
\begin{itemize}
\item
We propose \Name{}, the first automated approach to mitigating
compression side channels for arbitrary web content with security
guarantees.
\item
We implement \Name{} in a tool for PHP-based web server applications. 
\item
We demonstrate, through experiments, that \Name{} outperforms
state-of-the-art techniques in terms of compression performance and
security guarantees.
\end{itemize}


The rest of the paper is organized as follows.  First, we motivate our
work in Section~\ref{sec:mot} by explaining how compression side
channels are exploited.  Then, we review the DEFLATE compressor in
Section~\ref{sec:compressor} before presenting our enhancement.  Next,
we present our static analysis in Section~\ref{sec:taint} and code
instrumentation in Section~\ref{sec:codeInst}.  We present our
experimental results in Section~\ref{sec:experiment}, review the
related work in Section~\ref{sec:related}, and then give our
conclusions in Section~\ref{sec:conclusion}.

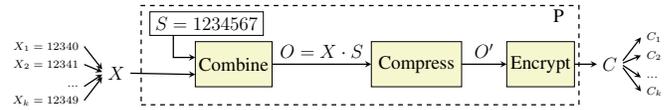
\begin{figure}
\vspace{1ex}

\centering
\scalebox{0.6}{
\begin{tikzpicture}[font=\scriptsize] 
  \tikzstyle{arrow}=[thick,->,>=stealth,black]
  \tikzstyle{txt}=[above=5pt,right,text width=1.6cm]
  \tikzstyle{long_txt}=[above=5pt,right,text width=2cm]
  \tikzstyle{short_txt}=[above=5pt,right,text width=1.4cm]
    	

  \node[align=center] (pinput) at (-0.5, -0.1) {\large $X$};

  \node[align=center, above left=1mm and 4mm of pinput] (pinput1) {$X_1 = 12340$};
  \node[align=center, above left=-3mm and 4mm of pinput] (pinput2) {$X_2 = 12341$};
  \node[align=center, above left=-7mm and 4mm of pinput] (pinput3) {$...$};
  \node[align=center, above left=-11mm and 4mm of pinput] (pinput4) {$X_k = 12349$};
 
  \node[draw=black] (sinput) at (1.5, 1.0) {\large $S=1234567$};
  \node[draw=black, fill=champagne, minimum width=17mm, minimum height=10mm, align=center, inner sep=0,outer sep=0] (append) at (2.1, 0.1) {\large Combine};
  
  \node[draw=black, fill=champagne, minimum width=20mm, minimum height=10mm, align=center, inner sep=0,outer sep=0, right=22mm of append] (compress) {\large Compress};

  \node[draw=black, fill=champagne, minimum width=15mm, minimum height=10mm, align=center, inner sep=0, outer sep=0, right=10mm of compress] (encrypt) {\large Encrypt};
  
  \node[align=center,right=5mm of encrypt] (output) {\large $C$};

  \node[align=center, above right=1mm and 4mm of output] (output1) {$C_1$};
  \node[align=center, above right=-3mm and 4mm of output] (output2) {$C_2$};
  \node[align=center, above right=-7mm and 4mm of output] (output3) {$...$};
  \node[align=center, above right=-11mm and 4mm of output] (output4) {$C_k$};
 
  \node[minimum width=97mm, minimum height=22mm, dashed, draw=black, thick, align=center] (P) at (4.9, 0.3) {};
  \node[text width=4.0cm, align=center, above=-5mm of P, xshift=44mm] {\large P};
  
  
   \draw [arrow] (pinput1.east) -- (pinput.165);
   \draw [arrow] (pinput2.east) -- (pinput.175);
   \draw [arrow] (pinput3.east) -- (pinput.185);
   \draw [arrow] (pinput4.east) -- (pinput.195);
   \draw [arrow] (pinput.east) -- (append.195); 
   \draw [arrow] (sinput.200) |- (append.170); 
   \draw [arrow] (append.east) -- (compress.west) node [midway, above, sloped] {\large $O = X \cdot S$};
   \draw [arrow] (compress.east) -- (encrypt.west) node [midway, above, sloped] {\large $O'$};
   \draw [arrow] (encrypt.east) -- (output.west);


   \draw [arrow] (output.15) -- (output1.west);
   \draw [arrow] (output.5) -- (output2.west);
   \draw [arrow] (output.355) -- (output3.west);
   \draw [arrow] (output.345) -- (output4.west);

\end{tikzpicture}
}
\vspace{-2ex}
\caption{The adversary model.}
\label{fig:adv-model}
\vspace{-2ex}
\end{figure}

\section{Motivation}
\label{sec:mot}

In this section, we explain what compression side-channels are and how
they are exploited.

\subsection{The Adversary Model}
\label{sec:adv_model}

First, we present an adversary model to illustrate the security risks.  
Suppose an attacker wishes to decrypt part of an encrypted message $C$
with some compressed plaintext $O$ produced by a procedure $P$, which
takes an input string $X$.  Fig.~\ref{fig:adv-model} shows the
scenario.  The attack can be achieved if the following conditions are met.
(1) $O$ contains some sensitive string $S$ of interest. 
(2) $O$ contains some other input string $X$.
(3) $P$ can be executed repeatedly with $X$ chosen by the attacker. 
(4) $P$ uses some form of dictionary compression on $O$.
(5) The size of the $C$ is visible to the attacker.

Conditions 1, 2 and 4 mean that $X$ and $S$ are always
compressed together.  Conditions 3 and 5 mean that the attacker
can observe how different the compressed sizes are when different
choices of $X$ are combined with $S$ to form $O$.  

As an example, consider the sensitive string $S$=\textcode{1234567}.  
When $X$ is \textcode{12344}, LZ77 will encode the first four
characters of $X$ with a reference; let this output be $C_1$.  When
$X$ is \textcode{12345}, the first five characters will be encoded;
let this output be $C_2$.
All other content being equal, $C_1$ will be one byte larger than
$C_2$ because of how $X$ is encoded.
In general, the size of $C$ becomes smaller as $X$ becomes
more \textit{similar} to $S$. This information can be exploited to
infer the content of $S$.

\subsection{A Realistic Attack Example}

We now show a scenario where the victim uses a webmail, and the
attacker wants to know who are in the victim's addressbook.  Since the
webmail uses the DEFLATE compression followed by encryption to process
data passing between the victim's browser and the webmail server, and
the attacker may observe the size of the data in
transition~\cite{Kelsey2002,Thai2012,Gluck:2,VanGoethem16Heist}, there
is a side-channel leak.
For example, if the attacker is on the same network as the victim and
tricks the victim into visiting a malicious web page, he may leverage
cookies in the browser to send requests to the server on behalf of the
victim.

\begin{figure}
\vspace{1ex}
\centering
\begin{minipage}{.95\linewidth} 
\include{figures/ex_php_code}
\end{minipage}
\caption{Compression side channel in a PHP program that creates an HTML page of a user's email addressbook.}
\label{fig:mvex1}
\end{figure}

Fig.~\ref{fig:mvex1} shows the server-side PHP code responding to such a request,
which generates the user's email addressbook on the returned HTML page.
In PHP, the \textcode{echo} statements (e.g., at line 17) cause data
to be compressed and then encrypted before they are sent to the
client.

The sensitive data are retrieved at line~10 using the API
function \textcode{get\_addressbook\_entries()}.
Each \textcode{\$entry} is also considered sensitive.
Following the data flow, we identify a sink of the tainted data at
line~17, which uses the function \textcode{htmlTag} (defined at lines
2-8) to construct a string and then feeds it to the \textcode{echo}.

Lines 23-25 print a \emph{hidden field} to the HTML, containing the
query string of the request URL.  Passing back the request URL in a
hidden field is common practice in server-generated HTML pages, e.g.,
to redirect back to that URL after a form's input is handled.
In this attack, the hidden field will be exploited by the attacker to
infer the sensitive data.

Now, we relate the example to the adversary model in
Section~\ref{sec:adv_model}: the PHP program is $P$, the email address
is the sensitive string $S$, the query string of request URL (hidden
field) is the input $X$, and the HTML is the output $O$.

The attacker causes the procedure $P$ to execute by making requests to
the server with the desired $X$ in the query string.  As mentioned
earlier, this is possible if the attacker makes requests from the
victim's browser, which has the victim's cookies and thus can make the
requests appear to be authenticated requests (however, the attacker
\textit{will not} see the content of the response HTML because of the 
\emph{same-origin} policy).  Nevertheless, the HTML file size is observable.

\begin{figure}
\vspace{1ex}
\centering
\scalebox{0.75}{
\begin{tabular}{l|c|c|l}
\hline
\textbf{Email addr.} & \multicolumn{2}{c|}{\textbf{Attacker's guesses}} & \textbf{Compressed data} \\
\hline\hline
sendto=bob@test.com & Iter. 1 & sendto=a &
sendto=bob@test.com\colorbox{yellowgreen}{(23,7)a}\\
\hline 
sendto=bob@test.com & Iter. 2 & sendto=b & sendto=bob@test.com\colorbox{yellowgreen}{(23,8)} \\
\hline
sendto=bob@test.com & Iter. 3 & sendto=c & sendto=bob@test.com\colorbox{yellowgreen}{(23,7)c} \\
\hline
\end{tabular}
}
\caption{Data containing a sensitive email address and three attacker's guesses lead to different compression sizes.}
\label{fig:mvdiagram}
\end{figure}

To understand how the attack works, consider
Fig.~\ref{fig:mvdiagram} as an example.  To decrypt an email
address, the attacker will attempt to guess it, character by character.
Assume that email addresses are composed from the alphabet:
$[A-Za-z0-9\_@.]$ and the addressbook contains an entry with the email
address
\textcode{bob@test.com}. 
For simplicity, also assume it is the only email address to be
displayed on the HTML page.

The attacker can \emph{bootstrap} with a known
prefix: \textcode{sendto=}.
To determine the next character, the attacker sends 64 requests to the
server, one for each guess character in the alphabet $[A-Za-z0-9\_@.]$
appended to the prefix.
The first response has
only \textcode{sendto=} compressed using the
reference \textcode{(23,7)}, whereas the second response has the
longer string \textcode{sendto=b} compressed using the
reference \textcode{(23,8)}.  With the other guess characters, the
compressed data size will be the same as that of
character \textcode{a}. Therefore, \textcode {b} is the correct guess.

On each iteration, the attacker determines one 
character. If the secret has $N$ characters, the attack takes as
little as $N$ iterations; this makes it extremely dangerous
in practice.

We choose to use the above example because of its simplicity
and ease of understanding, but more refined attacks also
exist~\cite{Gluck:2, Karakostas16}.  Nevertheless, the simple example
already illustrates the risk of the compression side channel.  Given
that 70\% of the top websites enable DEFLATE
compression~\cite{compStats}, this is particularly alarming to the
security-conscious.

\subsection{Our Mitigation Technique}

In \Name{}, we take a two-pronged approach.
First, we enhance the compressor to exercise the freedom to not
compress for any input data surrounded by special markers.
Second, we use static analysis and code instrumentation to identify
flow of sensitive data through the program and automatically
insert special markers.
In Fig.~\ref{fig:mvex1}, for example, it would identify the echo
of \textcode{htmlTag} at Line 17 as leaky.
Then, it would identify instrumentation points in the PHP code (sensitive
arguments \textcode{\$name} and \textcode{\$entry->email}
to \textcode{htmlTag}) for generating markers at run time.

Table~\ref{tbl:qualstats} compares \Name{} with state-of-the-art
techniques, which lack in performance, generality, or automation.

\begin{table}
\centering
\caption{Comparing our method to existing approaches.}
\label{tbl:qualstats}
\scalebox{0.8}{
\begin{tabular}{|l | c | c | c | c|} \hline
	Method & \makecell{Arbitrary\\ data?} & \makecell{Leak-free\\ Guarantee?} & \makecell{Fully\\ Automated?} & \makecell{High\\ Compression?} \\ \hline\hline

\textbf{\Name{}} (our new method)             & \textcolor{blue}{\checkmark} & \textcolor{blue}{\checkmark} & \textcolor{blue}{\checkmark} & \textcolor{blue}{\checkmark} \\ \hline
Keyword-based~\cite{SafeDeflate}              & \checkmark &            & \checkmark &            \\ \hline
Masking-based~\cite{Karakostas16CTX}          &            &  \checkmark    &  & \checkmark \\ \hline
Huffman-only~\cite{Huffman1952}               & \checkmark & \checkmark & \checkmark &            \\ \hline

\end{tabular}
}
\end{table}

\subsubsection{Keyword-based}

Techniques such as \emph{SafeDeflate}~\cite{SafeDeflate} utilizes two kinds
of keywords: a \emph{sensitive} alphabet ($A$) and a predefined
dictionary ($D$) of \emph{non-sensitive} strings.
%
%
%
Compression of a sequence of characters $L = l_0l_1...l_m$ is allowed
only if it does not begin or end with a sensitive character in $A$, or
it matches a string in $D$.  
%
%
%
%
This characterization of sensitive strings is unsound, and
provides no security guarantee. For example, a user may configure
the alphabet $A = [A-Za-z0-9\_@.]$ to protect emails, but the
match \textit{'=bob@test.com'} is still allowed under this
configuration. Furthermore, keyword-based techniques
degrade the runtime performance and 
compression ratio substantially. We shall demonstrate
both problems experimentally in Section~\ref{sec:experiment}.

\subsubsection{Masking-based}

Techniques such as CTX~\cite{Karakostas16CTX} generate a random and
reversible \emph{masking} operation and apply it to sensitive data 
before sending it to the client.  On the client side, special
JavaScript code must be used as well to undo the masking.  The masked
data must be enclosed in HTML \emph{$<$div$>$} tags with some unique
ID attribute.  This has two drawbacks.
First, it does not work for many applications, including our example
in Fig.~\ref{fig:mvex1} because inserting \emph{$<$div$>$}
around \emph{bob@test.com} in the URL would break the link.
Fig.~\ref{fig:mvex2} shows the example HTML with the email address
at Line~3 and hidden field at Line~5.
While one could mask the whole HTML tag at Line 3, it would break the
initial parsing of HTML.  This problem also applies to sensitive
data in JavaScript code.
%

\subsubsection{Huffman-Only}

The naive approach to mitigating compression side channels in DEFLATE
is to disable LZ77 and use Huffman coding only.  This is guaranteed to
be secure because the only information that may be deduced from the
compressed file would be the \emph{symbol distribution}, but so far,
no attacks have been reported to exploit the symbol distribution.
However, this approach has a high performance penalty because, even in
the best case, Huffman coding can only reduce a single byte's size by
62.5\% (and on average much lower), which is significantly worse than
enabling LZ77.

Compared to these existing techniques, \Name{} has the advantage of
protecting arbitrary web content \emph{soundly} (with a security guarantee)
and
\textit{more efficiently} (in terms of compression ratio) while
requiring no manual effort.  In Fig.~\ref{fig:mvex1}, it would only
identify the email addresses as sensitive, and all other web content
would be free for compression.

\begin{figure}
\hspace{-5ex}
\begin{minipage}{1.07\linewidth} 
\input{figures/ex_HTML_code}
\end{minipage}
\caption{HTML response with guess character \textcode{b} at line~4.}
\label{fig:mvex2}
\end{figure}

\section{Enhancing the Compressor}
\label{sec:compressor}

We first review the state-of-the-art DEFLATE compressor, and then
present our enhancement.

\subsection{The Original Compressor}
\label{sec:deflatealgo}

The DEFLATE compressor, shown in Algorithm~\ref{alg:deflatealgo}, combines
Huffman coding~\cite{Huffman1952} and LZ77 matching~\cite{lz77}, and
can be found in various open-source libraries
including \textit{zlib}~\cite{zlib}.

\subsubsection{Huffman Coding}
\label{sec:Huffman}

Given an input composed of symbols from some alphabet, it assigns bit
strings to symbols and then encodes the input using these bit
strings. It saves space by assigning shorter bit strings to symbols
used more frequently.
No bit string is a prefix of another bit string, which avoids
ambiguity when decoding bit strings back into symbols.
Fig.~\ref{fig:compex} shows an example of applying Huffman coding to
part of a popular tongue-twister.  There are two columns in the row of
Huffman coding: the right shows bit strings assigned to each input
symbol, based on the symbol count distribution, and the left shows the
encoding result.

\subsubsection{LZ77 Matching}
\label{sec:lz77}

In dictionary compression, a set of literal strings, called
a \emph{dictionary}, is stored in the compressed file, and any input
string that matches a dictionary string is encoded as
a \emph{reference}. In LZ77~\cite{lz77}, the dictionary is the input
itself, and each reference can only point to locations earlier in the
input.  The dictionary string is called a \emph{match} and the encoded
data, or reference, is called the \emph{reproducible extension}.
%
%
Fig.~\ref{fig:compex} illustrates LZ77 matching in the third
row, where the input has two redundant strings, \textcode{she}
and \textcode{lls}.  They are encoded as references $\langle
13,3\rangle$ and $\langle 10,3\rangle$, respectively.

\begin{figure}[h]
\centering
\scalebox{0.7}{

\begin{tabular}{|l | l | l|}
\hline
\multirow{2}{*}{Input string to be compressed} & \multicolumn{2}{c|}{\multirow{2}{*}{she sells seashells}} \\
                                           & \multicolumn{2}{c|}{}         \\\hline
\multirow{4}{*}{Output of Huffman-coding only } & \multicolumn{1}{c|}{Encoding} & Huffman Codes \\\cline{2-3}
     & 11 100 00 1011 11 00 & $e$ = 00, $h$ = 100 \\
      & 01 01 11 1011 11 00 & $l$ = 01, $a$ = 1010 \\
     & 1010 11 100 00 01 01 11 & $s$ = 11, \textquotesingle~\textquotesingle = 1011 \\    
	\hline

\multirow{2}{*}{Output of LZ77 matching only} & \multicolumn{2}{c|}{\multirow{2}{*}{she sells sea$<$13,3$><$10, 3$>$}} \\
    &  \multicolumn{2}{c|}{}  \\\hline
\end{tabular}
}
\caption{Example for Huffman coding  and LZ77 matching.}
\label{fig:compex}
\end{figure}


%
%
%

\begin{algorithm}[t]
\caption{The DEFLATE compression procedure.\label{alg:deflatealgo}}
{\footnotesize
\begin{algorithmic}[1]
\STATE \proc{DeflateCompress}($input$)~~\{
	\STATE \Ind{1} $buf \leftarrow$ \proc{LZ77Matching}($input$);
	\STATE \Ind{1} $output \leftarrow$ \proc{HuffmanCoding}($buf$);
	\STATE \Ind{1} \textbf{return} $output$;
%
\STATE \}
\end{algorithmic}
}
\end{algorithm}

\subsection{Our Modified Compressor}
\label{sec:new-compressor}

Our modified LZ77-matching procedure includes a pre-processing step,
which identifies and removes the annotations of sensitive data (in the
form of special tokens).  While removing these special tokens, it also
computes the necessary metadata that indicate the locations of the
sensitive data.  Viewing the compressor's input as an array of bytes,
the metadata stores, for each index in the input buffer, its forward
distance to the next closest region of sensitive data.

The procedure is shown in Algorithm~\ref{fig:debreachalgo}, where the
metadata \textcode{nextTaint[i]} denotes the distance from $ input[i] $ 
to the next region of sensitive data (i.e. $ input[i + nextTaint[i]] $ is sensitive).
\textcode{nextTaint[i]==0} means that
$input[i]$ is sensitive. The procedure incrementally and forwardly
considers each index $ i $ in the input beginning from $ i=0 $. When
the current index is not sensitive (lines 9-34), we first search the
previous input (i.e., $ input[0..i-1] $) for matches (lines 9-22). 
If a match is found, an LZ77 reference is output, and $ i $ is
incremented by the length of the match (lines 23-29). Otherwise, $ i $
is incremented by 1, and the literal input byte is output (lines
30-33). 

To quickly find matches, a dictionary is maintained that records the
positions of previously seen strings. The dictionary is updated with
every string in the input buffer (lines 25-27 and 31), unless $ i $ is
in a sensitive region (lines 5-8). The dictionary is then queried for
positions of candidates matches for the current index (line 10). For
compression ratio efficiency, a minimum match length is defined ($
minMatch = 3 $ for both \Name{} and the original zlib), and for memory
efficiency, only strings of length $ minMatch $ are stored in the
dictionary. However, when searching for a match, we compute the
maximum length (lines 12-16), so the best match is \textit{always}
found in the previous input buffer.

\begin{algorithm}[t]	
\caption{Our new LZ77 matching in \Name{}.}
\label{fig:debreachalgo}
{\footnotesize
\begin{algorithmic}[1]
\STATE \proc{LZ77Matching}($\mathit{input}$, \textcolor{dkblue}{$\mathit{nextTaint}$}) ~~\{
\STATE \Ind{1} initialize $\mathit{dict}, \mathit{output} $ to empty
\STATE \Ind{1} $i = 0$
\STATE \Ind{1} \textbf{while} ($i < \mathit{input.length}$) ~~\{
\STATE \Ind{2}    \textbf{if} (\textcolor{dkblue}{$\mathit{nextTaint}$}$[i] \leq minMatch$)~~\{  \textcolor{gray}{// skip sensitive data}
\STATE \Ind{3}       $\mathit{output} \mathrel{+}= \mathit{input[i]}$
\STATE \Ind{3}       $i \mathrel{+}= 1$
\STATE \Ind{2}    \} \textbf{else} ~~\{ \textcolor{gray}{// search for matches}
\STATE \Ind{3}       $\mathit{bestLen = bestDist = 0}$
\STATE \Ind{3}       \textbf{foreach} ($\mathit{matchLoc}$ in $\mathit{dict[input[i..i+minMatch]]}$) ~~\{
\STATE \Ind{4}         $\mathit{maxLen} = \mathit{min}$ (\textcolor{dkblue}{$\mathit{nextTaint}$}$[i]$, \textcolor{dkblue}{$\mathit{nextTaint}$}$[\mathit{matchLoc}]$)
\STATE \Ind{4}         $\mathit{matchLen} = 0 $
\STATE \Ind{4}         \textbf{while} ($ input[i+matchLen] ==$ 
\STATE \Ind{7}         $ \; input[matchLoc+matchLen] $) ~~\{
\STATE \Ind{5}         $ matchLen \mathrel{+}= 1 $
\STATE \Ind{4}         \}
\STATE \Ind{4}         $\mathit{len} = \mathit{min}~(\mathit{maxLen}, \mathit{matchLen})$
\STATE \Ind{4}         \textbf{if} ($\mathit{len} > \mathit{bestLen}$) ~~\{
\STATE \Ind{5}            $\mathit{bestLen} = \mathit{len}$
\STATE \Ind{5}            $\mathit{bestDist} = i - \mathit{matchLoc}$
\STATE \Ind{4}         \}
\STATE \Ind{3}       \}
\STATE \Ind{3}       \textbf{if} ($\mathit{bestLen > minMatch}$) ~~\{
\STATE \Ind{4}         $\mathit{output} \mathrel{+}= \langle \mathit{bestDist, bestLen} \rangle$
\STATE \Ind{4}         \textbf{foreach} ($ j \in i..i+bestLen-minMatch$) ~~\{
\STATE \Ind{5}            $ dict[input[j..j+minMatch]] = j $
\STATE \Ind{4}         \}
\STATE \Ind{4}         $i = i+\mathit{bestLen}$
\STATE \Ind{3}       \} \textbf{else} ~~\{
\STATE \Ind{4}         $\mathit{output}$ += $\mathit{input[i]}$
\STATE \Ind{4}         $ dict[input[i..i+minMatch]] = i $
\STATE \Ind{4}         $i \mathrel{+}= 1$
\STATE \Ind{3}       \}
\STATE \Ind{2}    \}
\STATE \Ind{1} \}
\STATE \Ind{1} \textbf{return} $\mathit{output}$
\STATE \}
\end{algorithmic}
}
\end{algorithm}

\subsubsection{On the correctness} 

\begin{figure}
	\centering
	
	\begin{subfigure}{.45\textwidth}
		\centering
		\scalebox{0.85}{\begin{tikzpicture}[square/.style={regular polygon,regular polygon sides=4, inner sep=0pt, outer sep=0pt, align=center}, font=\footnotesize]
  \tikzstyle{node}=[minimum size=0pt]
  \tikzstyle{nnode}=[minimum size=0pt,inner sep=0pt]
  \tikzstyle{lnode}=[circle,draw,minimum size=4pt,inner sep=0pt,fill]
 
  \node (r0) [draw, minimum width=80mm,minimum height=4mm]  at (0,0)  {};
 
  \node (r1) [draw, minimum width=10mm,minimum height=4mm, right=-75mm of r0] {};
  
  \node (r2) [draw,
            line space=8pt,  minimum width=10mm,minimum height=4mm, right=-35mm of r0] {};
            

  \node (r3) [draw, thick, color=red,  fill=red, fill opacity=0.2, minimum width=11mm, minimum height=4mm, right=-42mm of r0] {};
  

  \node[node] (c0)[left=3mm of r1, align=center, inner sep=0pt, outer sep=1pt, xshift=-5mm]  {\small Case 1};

  \node[node] (i0)[below=3mm of r1, align=center, inner sep=0pt, outer sep=1pt, xshift=-5mm]  {matchLoc};
  \draw[->,thick] (i0.north) to (r1.200);
  
  \node[node] (i1)[above=3mm of r1, align=center, inner sep=0pt, outer sep=1pt, xshift=5mm]  {matchLoc + len};
  \draw[->,thick] (i1.south) to (r1.20);
  
  \node[node] (i2)[below=3mm of r2, align=center, inner sep=0pt, outer sep=1pt, xshift=-12.5mm]  {i};
  \draw[->,thick] (i2.north) to (r2.188);

  \node[node] (i3)[above=3mm of r2.north east, align=center, inner sep=0pt, outer sep=1pt]  {i + len};
  \draw[->,thick] (i3.south) to (r2.north east);  
  
  

\end{tikzpicture}
\vspace{-2ex}}
	\end{subfigure}
	
	\vspace{2.5mm}
	
	\begin{subfigure}{.45\textwidth}
		\centering
		\scalebox{0.85}{\begin{tikzpicture}[square/.style={regular polygon,regular polygon sides=4, inner sep=0pt, outer sep=0pt, align=center}, font=\footnotesize]
  \tikzstyle{node}=[minimum size=0pt]
  \tikzstyle{nnode}=[minimum size=0pt,inner sep=0pt]
  \tikzstyle{lnode}=[circle,draw,minimum size=4pt,inner sep=0pt,fill]
  
  \node (r0) [draw, minimum width=80mm,minimum height=4mm]  at (0,0)  {};
  
  \node (r1) [draw, minimum width=10mm,minimum height=4mm, right=-75mm of r0] {};
  
  \node (r2) [draw,
            line space=8pt,  minimum width=10mm,minimum height=4mm, right=-35mm of r0] {};
            

  \node (r3) [draw, thick, color=red,  fill=red, fill opacity=0.2, minimum width=11mm, minimum height=4mm, right=-30mm of r0] {};
  
  
  \node[node] (c0)[left=3mm of r1, align=center, inner sep=0pt, outer sep=1pt, xshift=-5mm]  {\small Case 2};
  
  \node[node] (i0)[below=3mm of r1, align=center, inner sep=0pt, outer sep=1pt, xshift=-5mm]  {matchLoc};
  \draw[->,thick] (i0.north) to (r1.200);
  
  \node[node] (i1)[above=3mm of r1, align=center, inner sep=0pt, outer sep=1pt, xshift=5mm]  {matchLoc + len};
  \draw[->,thick] (i1.south) to (r1.20);
  
  \node[node] (i2)[below=3mm of r2, align=center, inner sep=0pt, outer sep=1pt, xshift=-12.5mm]  {i};
  \draw[->,thick] (i2.north) to (r2.188);

  \node[node] (i3)[above=3mm of r2.north east, align=center, inner sep=0pt, outer sep=1pt]  {i + len};
  \draw[->,thick] (i3.south) to (r2.north east);

\end{tikzpicture}
\vspace{-2ex}}
	\end{subfigure}
	
	\vspace{2.5mm}
	
	\begin{subfigure}{.45\textwidth}
		\centering
		\scalebox{0.85}{\begin{tikzpicture}[square/.style={regular polygon,regular polygon sides=4, inner sep=0pt, outer sep=0pt, align=center}, font=\footnotesize]
  \tikzstyle{node}=[minimum size=0pt]
  \tikzstyle{nnode}=[minimum size=0pt,inner sep=0pt]
  \tikzstyle{lnode}=[circle,draw,minimum size=4pt,inner sep=0pt,fill]
  
  \node (r0) [draw, minimum width=80mm,minimum height=4mm]  at (0,0)  {};
  
  \node (r1) [draw, minimum width=10mm,minimum height=4mm, right=-75mm of r0] {};
  
  \node (r2) [draw,
            line space=8pt,  minimum width=10mm,minimum height=4mm, right=-35mm of r0] {};
            

  \node (r3) [draw, thick, color=red,  fill=red, fill opacity=0.2, minimum width=11mm, minimum height=4mm, right=-78mm of r0] {};
  
  \node[node] (c0)[left=3mm of r1, align=center, inner sep=0pt, outer sep=1pt, xshift=-5mm]  {\small Case 3};

  \node[node] (i0)[below=3mm of r1, align=center, inner sep=0pt, outer sep=1pt, xshift=-5mm]  {matchLoc};
  \draw[->,thick] (i0.north) to (r1.200);
  
  \node[node] (i1)[above=3mm of r1, align=center, inner sep=0pt, outer sep=1pt, xshift=5mm]  {matchLoc + len};
  \draw[->,thick] (i1.south) to (r1.20);
  
  \node[node] (i2)[below=3mm of r2, align=center, inner sep=0pt, outer sep=1pt, xshift=-12.5mm]  {i};
  \draw[->,thick] (i2.north) to (r2.188);

  \node[node] (i3)[above=3mm of r2.north east, align=center, inner sep=0pt, outer sep=1pt]  {i + len};
  \draw[->,thick] (i3.south) to (r2.north east);

\end{tikzpicture}
\vspace{-2ex}}
	\end{subfigure}
	
	\vspace{2.5mm}
	
	\begin{subfigure}{.45\textwidth}
		\centering
		\scalebox{0.85}{\begin{tikzpicture}[square/.style={regular polygon,regular polygon sides=4, inner sep=0pt, outer sep=0pt, align=center}, font=\footnotesize]
  \tikzstyle{node}=[minimum size=0pt]
  \tikzstyle{nnode}=[minimum size=0pt,inner sep=0pt]
  \tikzstyle{lnode}=[circle,draw,minimum size=4pt,inner sep=0pt,fill]
  
  \node (r0) [draw, minimum width=80mm,minimum height=4mm]  at (0,0)  {};
  
  \node (r1) [draw, minimum width=10mm,minimum height=4mm, right=-75mm of r0] {};
  
  \node (r2) [draw,
            line space=8pt,  minimum width=10mm,minimum height=4mm, right=-35mm of r0] {};
            

  \node (r3) [draw, thick, color=red,  fill=red, fill opacity=0.2, minimum width=11mm, minimum height=4mm, right=-70mm of r0] {};
  
  \node[node] (c0)[left=3mm of r1, align=center, inner sep=0pt, outer sep=1pt, xshift=-5mm]  {\small Case 4};
  
  \node[node] (i0)[below=3mm of r1, align=center, inner sep=0pt, outer sep=1pt, xshift=-5mm]  {matchLoc};
  \draw[->,thick] (i0.north) to (r1.200);
  
  \node[node] (i1)[above=3mm of r1, align=center, inner sep=0pt, outer sep=1pt, xshift=5mm]  {matchLoc + len};
  \draw[->,thick] (i1.south) to (r1.20);
  
  \node[node] (i2)[below=3mm of r2, align=center, inner sep=0pt, outer sep=1pt, xshift=-12.5mm]  {i};
  \draw[->,thick] (i2.north) to (r2.188);

  \node[node] (i3)[above=3mm of r2.north east, align=center, inner sep=0pt, outer sep=1pt]  {i + len};
  \draw[->,thick] (i3.south) to (r2.north east);

\end{tikzpicture}
\vspace{-2ex}}
	\end{subfigure}
	
	
	\caption{LZ77: four possible cases for sensitive data.}
	\label{fig:all-cases}
	\vspace{-2ex}
\end{figure}
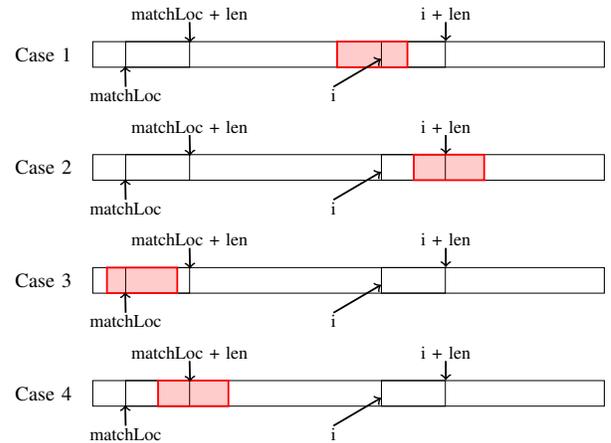

The procedure is correct in that LZ77 matches are not allowed with
sensitive data. To prove this, it suffices to show that neither of the
two components of an LZ77 match (the match and the reproducible
extension) may contain sensitive data. There are four cases to
consider where sensitive data may be included in a match, which are
illustrated in Fig.~\ref{fig:all-cases}. Each case represents a state
of Algorithm~\ref{fig:debreachalgo}, which consists of an input
buffer, a region of sensitive data shown in shaded red, a reproducible
extension ($ i $ and $ i + len $), and a match ($ matchLoc $ and $
matchLoc + len $).

First, observe that we never enter the match-searching block (lines
9-34) when $ i $ is sensitive (i.e., $ nextTaint[i] == 0 $) because of
the guard at line 5. This ensures that the reproducible extension
cannot start in sensitive data (case 1 in
Fig.~\ref{fig:all-cases}). Next, leveraging the fact that
$nextTaint[i] $ is the (forward) distance to the \textit{closest} region
of sensitive data (i.e., $ input[i] $ to $ input[i + nextTaint[i]-1]
$ is \emph{not} sensitive), it also cannot extend into sensitive data
(case 2). This is because, when searching for a match (lines 9-22),
we set a maximum match length (line 11), which is the minimum of $
nextTaint[i] $ and $ nextTaint[matchLoc] $. This guarantees the
reproducible extension cannot extend into sensitive data.

In addition, a match cannot start in sensitive data (case 3). Notice
that the dictionary is only updated at lines 25-27 and 31 with the
locations in the reproducible extension or the literal byte, which we
just showed cannot not contain sensitive data. Since we only search
the dictionary for match locations (line 10), the property then
follows. Finally, the match also cannot extend into sensitive data
(case 4) since we set a maximum match length (line 11).

The correctness then follows trivially, since the two components of an
LZ77 match cannot start in nor extend into sensitive data.

\subsubsection{On the security guarantee} 

Our enhanced compression algorithm guarantees freedom from
leaks \emph{due to LZ77} about the \emph{literal content} of the
demarcated sensitive data. This follows from the correctness of the
algorithm since LZ77 matches cannot contain sensitive data, which
implies there can be no dependency (in the behavior of LZ77) between
an attacker-controlled plaintext and the literal content of the
sensitive data. 

However, we do not attempt to eliminate other leaks about other information,
such as the length of the sensitive data. In addition, we do not
prevent purely size-based side channel leaks, e.g., as
in~\cite{zhang2010sidebuster}. This is analogous to how a mitigation
technique for power-based side channels does not mitigate leaks for
timing side channels.

Our algorithm guarantees a form
of \textit{non-interference}~\cite{almeida2016verifying,
sabelfeld2003language, goguen1982security} about the LZ77 matching
procedure. Informally, a prodedure guarantees non-interference if the
values of secrets do not influence its behavior in an observable
way. Formally stated, let $ P $ be our program (Algorithm~\ref{fig:debreachalgo}), with public string inputs
$ X = \{x_1,...,x_n\} $ and secret string inputs $ S = \{s_1,...,s_m\} $, i.e. $ P(X, S) $ is:
\[
 \textcode{LZ77Matching}(x_1 \cdot s_1 \cdot x_2 \cdot ... \cdot s_m \cdot x_n, ~ nextTaint)~,
\]

Then, let $ R_P(X, S) $ be the information observable to the attacker, which in our case is the length of the output of $ P $ (references have size 1). The
non-interference property that Algorithm~\ref{fig:debreachalgo} guarantees
is:
\begin{gather*} 
\forall X, S, S', \mathrm{such~that}~\forall i ~ |s_i| = |s'_i| . \\
R_P(X, S) = R_P(X, S')~.
\end{gather*}
The above property says that, for any fixed public inputs $ X $, changing the content of the secret variables does not affect the attacker's observations, however changing the length of the secrets may.


\section{Identifying the Sensitive Data}
\label{sec:taint}

We develop a conservative static analysis to identify the sensitive
data.
Results of the analysis are then used to perform code instrumentation,
to be presented in Section~\ref{sec:codeInst}.

\subsection{Sources of Sensitive Data}
\label{sec:sources}

Our analysis relies on a set of sensitive APIs that act as tainted
sources.  In Fig.~\ref{fig:mvex1}, for instance, the sensitive
function is \textcode{get\_addressbook\_entries()}.  For server
applications targeted by \Name{}, the choice of sensitive APIs is
often obvious.
For example, when the PHP code implements a webmail application,
sensitive data are the user's email data and the PHP program must
connect to an email provider's IMAP server to retrieve email data.
Therefore, the functions that communicate with the IMAP server are
sensitive.

The webmail application \textit{NOCC}~\cite{NOCC}, in particular, uses
PHP's IMAP module, and thus built-in functions such as
\textit{imap\_fetchbody} and \textit{ imap\_fetchheader} are treated 
as sensitive API functions.  The application
named \textit{Squirrelmail}~\cite{Squirrelmail}, on the other hand,
implements its own IMAP communication API; in this case, it uses two
functions named \textit{sqimap\_run\_command}
and \textit{sqimap\_run\_command\_list}, which are treated as
sensitive API functions.

Other server applications may store sensitive data in a database, for
which PHP provides API functions. Therefore, these APIs should be
treated as sensitive.
For example, the addressbook applications \textit{iAddressbook}~\cite{iAddressbook}
and \textit{Addressbook}~\cite{Addressbook} store sensitive contact information in
a database.

Database administration tools such as \textit{Adminer}~\cite{Adminer}
also have sensitive APIs.  
For example, \textit{Adminer}~\cite{Adminer} is designed with
the \textit{select} function used as an API for accessing row data.
Thus, for \Name{}, considering this as sensitive will achieve the
desired security and compression performance.

%
%
%

\subsection{Datalog-based Taint Analysis}

We follow a declarative program analysis framework~\cite{whaley2004,
lam2005,livshits2005, mangal2015}, which first traverses the PHP
code to construct a Datalog program, consisting of a set
of \textit{facts} and a set of \textit{inference rules}.  The Datalog
facts are relations that are known to hold in the PHP code.  The
inference rules define how to deduce new facts from existing facts.
Since the analysis can be formulated as a fixed-point computation, in
Datalog, the inference rules will be applied repeatedly until all
facts are deduced.
Consider the example Datalog program below:

\vspace{1ex}
{\footnotesize
\begin{tabular}{ll}
	\proc{Edge}$(n_1, n_2)$ & \\
	\proc{Edge}$(n_2, n_3)$ & \\
	\proc{Path}$(x, y)$     & $\leftarrow$ \proc{Edge}$(x, y)$ \\
	\proc{Path}$(x, y)$     & $\leftarrow$ \proc{Edge}$(x, z)$, \proc{Path}$(z, y)$
\end{tabular}
}
\vspace{1ex}

\noindent 
The first two lines define facts regarding nodes in a graph: there is
an edge from $n_1$ to $n_2$ and another edge from $n_2$ to $n_3$.  The
next two lines define the inference rules, saying that (1) if there is
an edge from $x$ to $y$, there is a path from $x$ to $y$; and
(2) \proc{Path} is transitively closed.
Any Datalog engine may be used then to solve the program, the
result of which is the set of all pairs that satisfy the \proc{Path}
relation.  By querying the result, we know if \proc{Path}$(n_1,n_3)$
holds.



\subsubsection{Generating Datalog Facts}

First, we construct an inter-procedural control flow graph (ICFG) from
the input program where each node in the ICFG corresponds to a program
statement. We perform flow- and context-insensitive analyses to
determine targets of method calls. 
%
We also hard-code data flow information for PHP built-in string functions, such
as \textcode{substr}, to make our analysis more accurate.

Our analysis handles arrays and fields. For arrays, we create
additional program variables for statically known indexes up to one
dimension, and we assume unknown indexes could refer to any of these
variables.  For fields, we take an
inexpensive \textit{object-insensitive} and
\textit{field-based} approach as proposed by Anderson~\cite{andersen1994};
it means we do not distinguish the same field names from different objects.
The field-based approach is particularly appropriate because,
in PHP applications, many classes are only instantiated once in
any execution (e.g., a database handle class), or
the same field of different heap objects holds similar data (e.g., an email class).



Next, we traverse the ICFG to generate Datalog facts shown as the
relations in Fig.~\ref{fig:dlrels}.
The domains are $V$, the set of variables or
objects, $S$, the set of statements, and $F$, the set of fields used
in a program.
We encode control flow as \proc{Edge} between statements, control
dependence as \proc{CtrlDep} between statements, and leakage-prone
branches as \proc{UnsafeBranch}.
The store of a variable or field of an object is encoded as \proc{StoreVar}
or \proc{StoreField}, and similarly for \proc{LoadVar} or \proc{LoadField}.
Finally, sensitive API calls and \textcode{echo} statements are
encoded as \proc{Source} and \proc{Sink}, respectively.




\begin{figure}[h]
\centering
\resizebox{\linewidth}{!}{
\begin{tabular}{ll}

\toprule
\proc{Edge}($s_{1} : S, s_{2} : S$)
& Control flow edge from statement $s_{2}$ to statement $s_{1}$
\\
\proc{UnsafeBranch}($s_{1} : S$)
& Branch statement $s_{1}$  may cause  implicit data flows
\\
\proc{CtrlDep}($s_{1} : S$, $s_{2} : S$)
& Statement $s_{1}$ is control dependent on $s_{2}$
\\
\proc{StoreVar}($v_{1} : V, s_{1} : S$) 
& Variable or Object $v_1$ is stored at statement $s_{1}$
\\
\proc{StoreField}($f_{1} : F, s_{1} : S$) 
& Field $f_1$ of an object is stored at statement $s_{1}$
\\
\proc{LoadVar}($v_{1} : V, s_{1} : S$) 
& Variable or Object $v_1$ is loaded at statement $s_{1}$
\\
\proc{LoadField}($f_{1} : F, s_{1} : S$) 
& Field $f_1$ of an object is loaded at statement $s_{1}$
\\
\proc{Source}($s_{1} : S$) 
& Source of sensitive functions at statement $s_{1}$
\\
\proc{Sink}($s_{1} : S$) 
& Sink (i.e., \textcode{echo}) is at statement $s_{1}$
\\
\bottomrule
\end{tabular}
}
\caption{Input relations for our analysis.}
\label{fig:dlrels}

\end{figure}



\subsubsection{Generating Inference Rules}

Our Datalog rules, shown in Fig.~\ref{fig:dlrules}, compute (1) the load of
tainted data at an \textcode{echo} and (2) dependencies for data
originating from sensitive APIs.

\begin{figure}[h]
{
\centering
\resizebox{\linewidth}{!}{
\begin{tabular}{lcl}

\toprule
\proc{TaintedVarFrom}($v_1$, $s_1$, $s_1$) & $\leftarrow$
& \proc{StoreVar}($v_1$, $s_1$) $\wedge$ \proc{Source}($s_1$)
\\
\proc{TaintedVarFrom}($v_1$, $s_1$, $s_3$) & $\leftarrow$
& \proc{TaintedVarFrom}($v_1$, $s_1$, $s_2$) $\wedge$ \proc{Edge}($s_2$, $s_3$) \\
&& $\wedge$ $\neg$ \proc{StoreVar}($v_1$, $s_2$)
\\
\proc{TaintedFieldFrom}($f_1$, $s_1$, $s_1$) & $\leftarrow$
& \proc{StoreField}($f_1$, $s_1$) $\wedge$ \proc{Source}($s_1$)
\\
\proc{TaintedFieldFrom}($f_1$, $s_1$, $s_3$) & $\leftarrow$
& \proc{TaintedFieldFrom}($f_1$, $s_1$, $s_2$) $\wedge$ \proc{Edge}($s_2$, $s_3$)
\\
\hline
\proc{Tainted}($s_2$) & $\leftarrow$
& \proc{TaintedVarFrom}($v_1$, $s_1$, $s_2$) $\wedge$  \proc{LoadVar}($v_1$, $s_2$)
\\
\proc{Tainted}($s_2$) & $\leftarrow$
& \proc{TaintedFieldFrom}($f_1$, $s_1$, $s_2$) $\wedge$  \proc{LoadField}($f_1$, $s_2$)
\\
\proc{Tainted}($s_3$) & $\leftarrow$
& \proc{TaintedVarFrom}($v_1$, $s_1$, $s_2$) $\wedge$  \proc{LoadVar}($v_1$, $s_2$) \\
&& \proc{UnsafeBranch}($s_2$) $\wedge$ \proc{CtrlDep}($s_3$, $s_2$)
\\
\proc{Tainted}($s_3$) & $\leftarrow$
& \proc{TaintedFieldFrom}($f_1$, $s_1$, $s_2$) $\wedge$  \proc{LoadField}($f_1$, $s_2$) \\
&& \proc{UnsafeBranch}($s_2$) $\wedge$ \proc{CtrlDep}($s_3$, $s_2$)
\\
\proc{TaintedVarFrom}($v_1$, $s_1$, $s_1$) & $\leftarrow$
& \proc{Tainted}($s_1$) $\wedge$  \proc{StoreVar}($v_1$, $s_1$)
\\
\proc{TaintedFieldFrom}($f_1$, $s_1$, $s_1$) & $\leftarrow$
& \proc{Tainted}($s_1$) $\wedge$  \proc{StoreField}($f_1$, $s_1$)
\\
\hline
\proc{DataDep}($s_1$, $s_2$) & $\leftarrow$
& \proc{TaintedVarFrom}($v_1$, $s_1$, $s_2$) $\wedge$ \proc{LoadVar}($v_1$, $s_1$)
\\
\proc{DataDep}($s_1$, $s_2$) & $\leftarrow$
& \proc{TaintedFieldFrom}($f_1$, $s_1$, $s_2$) $\wedge$ \proc{LoadVar}($f_1$, $s_1$)
\\
\proc{TaintedSink}($s_1$) & $\leftarrow$
& \proc{Sink}($s_1$) $\wedge$ \proc{Tainted}($s_1$)
\\
\bottomrule
\end{tabular}
}
}
\caption{Datalog rules for data dependency analysis.}
\label{fig:dlrules}
\end{figure}

Let us walk through the rules in Fig.~\ref{fig:dlrules} to better
understand our approach.
The first and third rules capture when a variable or field is
assigned at a sensitive source statement.
The relation \proc{TaintedVarFrom}($v_1, s_1, s_2$) means 
$v_1$ defined in statement $s_1$ still holds sensitive data at statement $s_2$, and
similarly for \proc{TaintedFieldFrom}. These two rules
initialize the relation for all stored variables/fields at source statements.

The second and fourth rules propagate these relations through control
flow graph edges.  The difference is that \proc{TaintedVarFrom} can be
blocked by a new assignment (\proc{StoreVar}),
but \proc{TaintedFieldFrom} cannot be blocked by a new assignment of
the field because we do not discriminate different objects for the
same field name.

The fifth through eighth rules initialize \proc{Tainted}($s_1$) when
tainted data is used either explicitly or implicitly, meaning $s_1$
loads tainted variables. Then, the ninth and tenth rules create
new \proc{TaintedVarFrom} and \proc{TaintedFieldFrom} relations for
the stored variables at tainted statements.
\textcolor{black}{
The next two rules infer \proc{DataDep}, where 
\proc{DataDep}($s_1$, $s_2$) means  $s_1$ data-depends 
on $s_2$, which occurs when tainted data is propagated from $s_1$ to
$s_2$ and loaded at $s_2$.  }
Finally, \proc{TaintedSink} represents when tainted data is used at a \proc{Sink} statement,
i.e., it may be sent through an \textcode{echo} statement.

%
%
%
%
Our approach shares the limitations of other static analysis tools for
dynamically typed languages such as PHP, e.g., unsoundness in the
presence reflection~\cite{LivshitsSSLACGKMV15}. However, for the
benchmarks used in our experiments, we have confirmed such language
features did not affect soundness of our analysis.

\subsection{Implicit Flows}
\label{sec:implicit-flows}

The key to high compression performance is proper tracking of implicit data
flows.
Since web applications are string-building programs at their core,
tainted variables are frequently used in branch conditions. However,
naively tracking all these implicit flows would result in over-tainting
and low compression performance.
Instead, we develop sufficient conditions under which implicit flows
can be ignored. 
Our conditions label a branch as safe if all of the atoms of its
predicate are:
\begin{itemize}
	\item a variable (e.g., \textcode{if (\$var)}); 
	\item comparing a variable to a hard-coded value;
	\item comparing to the length of variable; or
	\item checking the type of a variable.
\end{itemize}
We do not attempt to protect Boolean variables from implicit data
flows, for two reasons.  First, even if we protect them during
compression, they may still be revealed through purely size-based side
channels (i.e., not due to compression).  Second, protecting them would
prevent us from ignoring frequently-occurring, performance-critical
branches that do not affect our security guarantee (specifically the
first criterion above). 


\subsection{Security Guarantee}
\label{sec:guarantees}

The guarantee we aim to provide is that an attacker cannot
discover \textit{literal} content of secret \textit{string type} data.
For example, the branch at line 13 in Fig.~\ref{fig:mvex1} can be
ignored, since it may only reveal that the length of the secret was
greater than 20 characters. 
Conversely, the case we \textit{do} care about is when a dynamically
determined non-sensitive string (not originating from a source) is
compared with a sensitive string and then eventually compressed, for
example:

\begin{figure}[H]
\vspace{-2ex}
\centering
\begin{minipage}{0.9\linewidth}
\begin{lstlisting}
$tainted = sensitive_source();
$untainted = $_GET["cgi_param"];
if ($tainted == $untainted)
	echo $untainted
\end{lstlisting}
\end{minipage}
\vspace{-3ex}
\end{figure}

Our analysis provides the security guarantee described above given two
assumptions. First, we assume that the truth value of dangerous
predicates such as the one above do not become associated with the
branches we ignore.  For example, we assume a dangerous predicate is
not first assigned to a variable, and subsequently used in a
branch. Second, we assume tainted data flows do not depend on dynamic
features such as reflection.  
%

\section{Instrumenting the Server Program}
\label{sec:codeInst}

We instrument the program to allow it to generate annotations of
sensitive data at run time, prior to the compression.

\subsection{Annotations}

Annotations of sensitive data are special markers inserted into the
tainted string value.  During the execution, the program randomly
generates a nonce of arbitrary length, and uses it to enclose the
sensitive string. For instance, in Fig.~\ref{fig:mvex1}, it would
wrap the (sensitive) argument at line 18 with
markers, e.g.,  \textcode{"DBR\{" . \$name . "\}DBR"}, where \textcode{DBR} is
the nonce.


\subsection{Efficient Code Instrumentation}

Naively, we could annotate all tainted variables used at
an \textcode{echo}, however this may result in the amount of annotated
data to become unacceptably large and degrade the compression ratio,
e.g., in Fig.~\ref{fig:mvex1}, where sensitive data
in \textcode{\$entry} are combined with non-sensitive HTML tags.

Naively annotating the tainted source is also problematic because
markers inserted into the string may affect subsequent manipulations,
e.g., in Fig.~\ref{fig:mvex1}, where an entry's name is truncated if
it is longer than 20 characters (lines 13-16) using
the \textcode{strlen} and \textcode{substr} operations.  Clearly,
inserting markers before these function calls changes their semantics.

To avoid both problems, we perform an analysis to determine the best
instrumentation point. To this end, we consider
both \emph{performance}, i.e., how much compression to maintain,
and \emph{safety}, i.e., not breaking the program semantics.

\begin{algorithm}[t]
{\footnotesize
	\caption{Computing the instrumentation points\label{fig:instralgo}}


	\begin{algorithmic}[1]
		\STATE \proc{InstrAnalysis}($tsinks$, $DDG$)~~\{
		\STATE \Ind{1} $instr\_pts =$ $[~]$
		\STATE \Ind{1} \textbf{for each} ($sink \in tsinks$) \{
		\STATE \Ind{2}     $ctx =$ function of $sink$
		\STATE \Ind{2}     $instr\_pts \; +=$ \proc{FindInstr}($sink$, $[~]$, $ctx$, $DDG$)
                \STATE \Ind{1} \}
		\STATE \Ind{1} \textbf{return} $instr\_pts$
                \STATE \}
	\end{algorithmic}

	\begin{algorithmic}[1]
		\STATE \proc{FindInstr}($cur$, $visited$, $ctx$, $DDG$) \{
		\STATE \Ind{1} $preds =$ immediate predecessors of $cur$ in $DDG$
		\STATE \Ind{1} \textbf{if} ($preds$ == $[~]$) 
		\STATE \Ind{2}     \textbf{return} $[cur]$
		\STATE \Ind{1} $preds =$ $\{p \in preds | p \not\in visited\}$
		\STATE \Ind{1} $visited \; +=$ $preds$
		\STATE \Ind{1} \textbf{if} ($preds$ == $[~]$)
		\STATE \Ind{2}     \textbf{return} $[~]$
		\STATE \Ind{1} \textbf{else if} (for all $p \in preds$, \proc{IsSafe}($p$, $DDG$) is $true$) \{
		\STATE \Ind{2}     $instr\_pts =$ $[~]$
		\STATE \Ind{2}     \textbf{for each} ($p \in preds$)
		\STATE \Ind{3}         $instr\_pts \; +=$ \proc{FindInstr}($p$, $visited$, $ctx$)
		\STATE \Ind{2}     \textbf{if} (any $pt \in instr\_pts$ is not in $ctx$)
		\STATE \Ind{3}         \textbf{return} $[cur]$
		\STATE \Ind{2}     \textbf{else}
		\STATE \Ind{3}         \textbf{return} $instr\_pts$
		\STATE \Ind{1} \}
		\STATE \Ind{1} \textbf{else}
		\STATE \Ind{2}     \textbf{return} $[cur]$
		\STATE \}
	\end{algorithmic}

}
\end{algorithm}

Our analysis in Algorithm~\ref{fig:instralgo} takes as input the
tainted echos and the data dependence graph (DDG) computed earlier.  On each tainted
echo, \proc{FindInstr} is called to find a set of statements (and
variables) to instrument. \proc{FindInstr} performs a backward search
along data dependence edges until it encounters a stopping condition,
which indicates the current statement is an instrumentation point.

There are three stopping conditions, namely: (a) the current statement
has no predecessors, indicating we reached the taint source (line 3),
(b) we have already visited all of the statement's predecessors,
indicating we have \emph{covered} it with another instrumentation
point (lines 6-7), or (c) some predecessor $P$ is unsafe (line 9).

A statement is unsafe if one of two conditions hold: (1) it contains
an operation that may be broken by inserted annotation, or (2) it
may affect another statement that meets condition (1). 
%
%
The check for condition (1) relies on a whitelisting based approach,
where all statements on the whitelist are guaranteed to not be affected by
inserting an annotation.  Generally speaking, condition (1) will \emph{not} be
satisfied (i.e., the statement is safe) if a statement consists of only assignment, concatenation,
or PHP built-in functions that can never be affected by inserted
annotation.

In Fig.~\ref{fig:mvex1}, statements in \textcode{htmlTag} are
safe because they only contain string concatenations, but lines 13
and 14 are not safe because annotations may break the called
functions. In addition, line 11 is not safe because 13 and 14 depend
on it.

\begin{figure}
\vspace{1ex}
\centering
\scalebox{0.7}{
	\begin{tikzpicture}
		\node[shape=circle,draw=black, dashed] (17) at (0,0) {17};
		\node[shape=circle,draw=black] (7) at (-1,0) {7};
		\node[shape=circle,draw=black] (6) at (-2,0) {6};
		\node[shape=circle,draw=black, thick, fill=blue!30] (18) at (-3,1) {18};
		\node[shape=circle,draw=black, fill=red!60] (14) at (-4,1) {14};
		\node[shape=circle,draw=black] (16) at (-4,2) {16};
		\node[shape=circle,draw=black] (5) at (-3,0) {5};
		\node[shape=circle,draw=black, thick, fill=blue!30] (20) at (-4,0) {20};
		\node[shape=circle,draw=black, fill=red!60] (11) at (-5,0) {11};

		\path[->] (7) edge node[left] {} (17);
		\path[->] (6) edge node[left] {} (7);
		\path[->] (5) edge node[left] {} (6);
		\path[->] (20) edge node[left] {} (5);
		\path[->] (11) edge node[left] {} (20);

		\path[->] (18) edge node[left] {} (6);
		\path[->] (14) edge node[left] {} (18);
		\path[->] (16) edge node[left] {} (18);
	\end{tikzpicture}
}

\caption{Data dependence graph of motivating example.}
\label{fig:ddg}
\vspace{-2ex}
\end{figure}
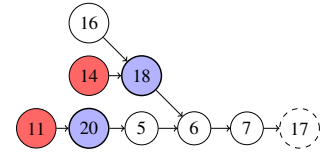

Our analysis also limits itself to annotating within the context
(i.e., a function or top-level script code) of the tainted sink (lines
13-16). This is to reduce over-annotating.
For example, real applications make use of utility functions for 
generating HTML, similar to the \textcode{htmlTag} function in Fig.~\ref{fig:mvex1}, and
they are used frequently with both sensitive and non-sensitive data.
Annotating inside these functions would likely degrade compression performance.

We now return to the example in Fig.~\ref{fig:mvex1}, whose DDG is
shown in Fig.~\ref{fig:ddg}.  
Nodes are labeled by line number, the
unsafe nodes for instrumentation are filled with red, and the tainted
echos are shown with a dashed circle.
%
During code instrumentation, we would step backward from node 17 and
branch at node 6. While considering node 18 along the top branch, the
safety check at line 9 of \proc{FindInstr} would fail because one of
its predecessors (node 14) is unsafe. When we reach node 20 along the
bottom branch, the safety check would fail again because node 11 is
unsafe. So we would annotate the variables at line 18 and 20 (blue
nodes in the graph).

\section{Experiments}
\label{sec:experiment}

We have implemented \Name{} using a combination of Java, Python, and
Datalog.  It requires the user to provide the top-level directory of the
PHP code and a set of sensitive API functions, and then generates
instrumented PHP code.
We leverage \textit{joern-php}~\cite{joern} to extract control-flow
and def/use Datalog facts from the PHP code.  We
extend \textit{joern-php} to handle arrays, fields, objects, globals,
and pass-by-reference parameters.
%
%
Our static analysis is implemented using 2K lines of Java code, 1.6K
lines of Python code, which includes our Datalog solver implemented in Python.

We conducted experiments to answer three questions:
\begin{itemize}
\item 
Are the static analysis and instrumentation components in \Name{}
efficient in handling real web applications?
\item 
Can \Name{} achieve high compression performance? In particular, how does it compare
to state-of-the-art techniques?
\item 
Can \Name{} eliminate the actual side channels? 
\end{itemize}
We answer the first two questions by comparing the compression ratios
of \Name{} to the keyword-based 
\emph{SafeDeflate}~\cite{SafeDeflate}, \emph{Huffman-Only}~\cite{Huffman1952}, and
an \textit{Oracle} version of \Name{}, which represents the practical
limit of \Name{}.

To produce the \emph{Oracle} version of \Name{}, we first instrument the application
with \Name{} and then manually inspect each instrumentation point to
decide if it can be optimized.  If, for example, the instrumentation
was due to a false-positive tainting, we remove it.  If it was a
true-positive, but a better instrumentation exists (that reduces the
amount of tainted data), we change it to the better one.

As for the third research question, we note that \Name{} guarantees to
eliminate the compression side channel.  Nevertheless, it is still informative to
demonstrate on real applications.  Thus, first, we show that leaks
indeed exist and can be exploited for some applications.  Then, we
show that leaks no longer exist in \Name{}-instrumented versions.

\subsection{Experimental Setup}

Our benchmark consists of five server applications with 145K lines of
PHP code in total.  The applications fall into three categories:
webmail, database administration, and addressbook.  Our main selection
criteria is that \Name{} has a practical use case for the application. The
characteristics are summarized in Table~\ref{tbl:appstats}, where
Columns 1 and 2 show the name and number of lines of code.  For each
application, Column 3 shows the five pages chosen for experiment; they
are web pages that users are likely to visit.  Column 4 shows the size
of the corresponding HTML response.
Response size can vary because of the application's
dynamic content. For example, email bodies can vary widely
in size.

\begin{table}
\caption{Characteristics of the benchmark applications.}
\label{tbl:appstats}
\centering
\scalebox{0.82}{
\begin{tabular}{|l|c|c|c|}\hline
Application  & Lines of Code &  Requested Page & Response HTML Size (KB)\\
& & &  \\\hline\hline
\multirow{5}{*}{Squirrelmail~\cite{Squirrelmail}} 
                              & \multirow{5}{*}{55,698} & compose email & 3       \\ \cline{3-4}
                              &                         & login         & 2       \\ \cline{3-4}
                              &                         & preferences   & 5       \\ \cline{3-4}
	                      &                         & view email    & 9 - 29  \\ \cline{3-4}
                              &	                        & view inbox    & 18 - 19 \\ \hline
\multirow{5}{*}{NOCC~\cite{NOCC}}         
                              & \multirow{5}{*}{17,610}	& compose email & 8       \\ \cline{3-4}
                              &                         & view email    & 24 - 102\\ \cline{3-4}
	                      &                         & view inbox    & 35 - 37 \\ \cline{3-4}
	                      &                         & login         & 6       \\ \cline{3-4}
	                      &                         & preferences   & 16      \\ \hline
\multirow{5}{*}{Adminer~\cite{Adminer}}      
                              &	\multirow{5}{*}{37,330}  & edit row      & 2 - 5   \\ \cline{3-4}
	                      &                         & insert item   & 2 - 4   \\ \cline{3-4}
	                      &                         & login         & 2       \\ \cline{3-4}
	                      &                         & table data    & 11 - 66 \\ \cline{3-4}
	                      &                         & table structure & 6 - 7 \\ \hline
\multirow{5}{*}{iAddressbook~\cite{iAddressbook}} 
                              &	\multirow{5}{*}{16,690}	& addressbook     & 14      \\ \cline{3-4}
	                      &                         & contact info    & 23 - 24 \\ \cline{3-4}
	                      &                         & edit contact    & 48      \\ \cline{3-4}
	                      &                         & login           & 2       \\ \cline{3-4}
	                      &                         & new contact     & 33      \\ \hline
\multirow{5}{*}{Addressbook~\cite{Addressbook}}  
                              &	\multirow{5}{*}{17,907}	& contact info    & 5       \\ \cline{3-4}
	                      &                         & edit contact    & 11      \\ \cline{3-4}
	                      &                         & login           & 4       \\ \cline{3-4}
	                      &                         & new contact     & 3       \\ \cline{3-4}
	                      &                         & addressbook     & 556     \\ \hline
\end{tabular}
}
\end{table}

To compare against the keyword-based \textit{SafeDeflate},
we need to configure its required \textit{sensitive alphabet} and the
dictionary of allowed strings, as described in the original
publication~\cite{SafeDeflate}.
For \emph{Huffman-Only}, we used the algorithm implemented
in \textit{zlib}~\cite{zlib}.
%

Next, we describe the applications in detail, and justify how we
configure and compare the different tools.


\subsubsection{Squirrelmail and NOCC}

These are webmail applications, where the user may want to prevent
email data from being leaked through the compression side channel.
In both applications, email data is retrieved from an IMAP server, so
for \Name{}, we choose functions that communicate with the IMAP server
as sensitive API functions.
%
%
For \textit{SafeDeflate}'s alphabet, we include alphanumerics, space,
comma, and period to protect natural language in email bodies, and we
include '/', ':', '\%', '\&', '=', and '?'  to protect links. In
addition, NOCC supports HTML in emails, so we add in '$<$', '$>$',
double quote, and single quote.  \emph{Squirrelmail} does not support
HTML, so we exclude these characters. For email addresses we include
'@' and '\_'.  While the alphabet provides protection in \textit{most}
cases, email bodies allow nearly any printable ASCII character,
so \emph{SafeDeflate} may still have leaks.


\subsubsection{Adminer}

This is a database administration (DBA) tool, where the user may want
to protect the rows of the database tables from being leaked.
%
%
Since \textit{Adminer} is designed with a function \textit{select}
that is used as an API for accessing row data, for \Name{}, we mark
this as the sensitive API.
We populate the database with a standard SQL dataset that resembles an
employee database for a company. The table's data is made of
alphanumerics and the hyphen character, so we
configure \textit{SafeDeflate} with this alphabet.


\subsubsection{Addressbook and iAddressbook}

These are open-source addressbook applications, where the user may
want to prevent the stored contact information from being leaked.
Both applications use a database for storing this information, so we
taint the query functions for both cases.
We populate the addressbook with ~500 contacts including names,
emails, phone numbers, addresses, jobs, and titles generated using
the \textit{faker} Node.js library. Accordingly, we configure
\emph{SafeDeflate} with alphanumerics, '-', '@', '.', and '\_'.

\subsection{Results of Analysis and Instrumentation}
\label{sec:results-analysis-instrumentation}

First, we measure the performance of \Name{}'s static analysis and
code instrumentation components.  The experiments were conducted on an
Ubuntu 16.04 machine with 12GB of RAM and an Intel i7 processor.

Table~\ref{tbl:analysis} shows the results.  Columns~1-3 show the
application name, page used in experiment, and total number of echos
involved.  Columns~4-5 show the number of tainted echos and
instrumentation points for \emph{oracle}.  Columns~6-8 show the
results of \Name{}, including the number of tainted echos, the number
of instrumentation points, and the analysis time.

\begin{table}
\caption{Performance of \Name{}'s analysis and instrumentation. For NOCC, etc., all five pages are the same (*).}
\label{tbl:analysis}
\centering
\scalebox{0.82}{
\begin{tabular}{|l|c|c|c|c|c|c|c|}\hline

\multirow{2}{*}{App} & \multirow{2}{*}{Page} & \multirow{2}{*}{\makecell{Total\\ Echos}} & \multicolumn{2}{c|}{Oracle} & \multicolumn{3}{c|}{\Name{} (new)} \\\cline{4-8}

 & & &\makecell{Tainted\\ Echos} &\makecell{Instr.\\ Points}
 &\makecell{Tainted\\ Echos} &\makecell{Instr.\\ Points} & Time
 (s) \\\hline

\multirow{5}{*}{Squirrelmail} &  comp-e & 96   & 10 & 15  & 15 & 17  & 159 \\
                              &  login  & 17   & 0  & 0   & 0  & 0   & 152 \\
                              &  pref.  & 58   & 13  & 12   & 13  & 12   & 151 \\
                              &  view-e & 85   & 6  & 12  & 7  & 9  & 156 \\
                              &  view-i & 99   & 12 & 36  & 21 & 37  & 152 \\\hline
\multirow{1}{*}{NOCC}         &  * & 2423 & 97 & 107 & 112 & 114 & 167 \\\hline
\multirow{1}{*}{Adminer}      &  * & 809  & 26 & 44  & 30 & 48  & 99  \\\hline
\multirow{1}{*}{iAddressbook} &  * & 763  & 83 & 104 & 84 & 106 & 129 \\\hline
\multirow{5}{*}{Addressbook}  &  addr-b & 220  & 30 & 59  & 31 & 65  & 78 \\
                              &  c-info & 228  & 14 & 19  & 15 & 19  & 72 \\
                              &  edit-c & 521  & 40 & 43  & 41 & 44  & 69 \\
                              &  login  & 220  & 30 & 59  & 31 & 65  & 78 \\
                              &  new-c  & 521  & 40 & 43  & 41 & 44  & 69 \\\hline
\end{tabular}
}
\end{table}

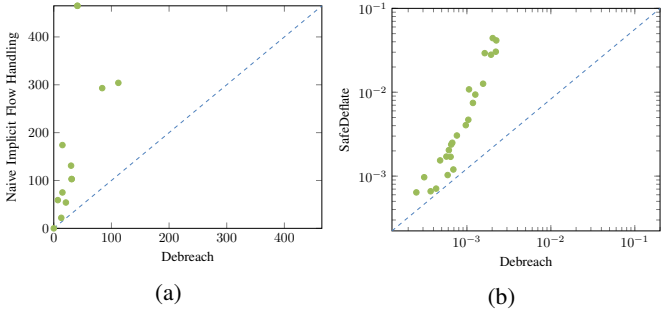
\begin{figure}
	\centering
	\begin{subfigure}{0.5\linewidth}
	\scalebox{0.52}{\begin{tikzpicture}
\begin{axis}[
        xmin=0,
        ymin=0,
        xmax=465,
        ymax=465,
        ylabel=Naive Implicit Flow Handling,
        xlabel=Debreach
    ]
    \addplot[
        scatter,
        only marks,
        point meta=explicit symbolic,
        scatter/classes={
            imp_flow={ggreen}
        }
    ] table [meta=label] {
x   y label
7    59  imp_flow
21   54   imp_flow
15   75   imp_flow
13   22   imp_flow
0    0  imp_flow
112  304  imp_flow
30   131  imp_flow
84   293  imp_flow
41   465  imp_flow
41   465  imp_flow
31   103  imp_flow
31   103  imp_flow
15   174  imp_flow
    };
    \draw [bblue,dashed] (rel axis cs:0,0) -- (rel axis cs:1,1);
\end{axis}
\end{tikzpicture}}
	\caption{\label{fig:imp_flow_comp}}
	\end{subfigure}%
	\begin{subfigure}{0.5\linewidth}
	\scalebox{0.52}{    \begin{tikzpicture}
\begin{axis}[
        xmin=0,
        ymin=0,
        xmax=0.2,
        ymax=0.1,
        axis equal,
        xmode=log,
        ylabel=SafeDeflate,
        ymode=log,
        xlabel=Debreach
    ]
    \addplot[
        scatter,
        only marks,
        point meta=explicit symbolic,
        scatter/classes={
            sdexec={ggreen}
        }
    ] table [meta=label] {
x   y label
0.01038	0.43913	sdexec
0.00069	0.00120	sdexec
0.00097	0.00405	sdexec
0.00059	0.00103	sdexec
0.00104	0.00470	sdexec
0.00065	0.00238	sdexec
0.00061	0.00204	sdexec
0.00037	0.00066	sdexec
0.00203	0.04420	sdexec
0.00057	0.00171	sdexec
0.00126	0.00938	sdexec
0.00156	0.01264	sdexec
0.00221	0.03044	sdexec
0.00031	0.00097	sdexec
0.00163	0.02924	sdexec
0.00025	0.00064	sdexec
0.00224	0.04148	sdexec
0.00194	0.02802	sdexec
0.00067	0.00250	sdexec
0.00118	0.00746	sdexec
0.00064	0.00170	sdexec
0.00043	0.00071	sdexec
0.00048	0.00154	sdexec
0.00076	0.00305	sdexec
0.00106	0.01082	sdexec

    };
    
    \draw [bblue,dashed] (rel axis cs:0,0) -- (rel axis cs:1,1);
\end{axis}
\end{tikzpicture}}
	\caption{\label{fig:sd_exec}}
	\end{subfigure}
	\caption{(a) Tainted echos with and without \Name{}'s implicit flow. (b) Execution time (s): \Name{} vs. \emph{SafeDeflate}.}
\end{figure}

The results show that \Name{} is close to \textit{oracle} in
identifying tainted echos and instrumentation points.  For most pages
there is $<$15\% false-positive rate in tainted echos. The majority of
false positives are due to implicit data flows.  A common case we
observe are branches that compare sensitive data to some configuration
parameter pulled from a source that we cannot prove to be
non-sensitive.  In addition, we show in Fig.~\ref{fig:imp_flow_comp}
the benefit of our implicit flow rules
(Section~\ref{sec:implicit-flows}), without which there would be 100's
of false positives. 

As for the time taken to complete the analysis and instrumentation,
Column~8 in Table~\ref{tbl:analysis} shows that \Name{} takes $<3$
minutes in all cases. Furthermore, it is scalable in handling real
applications.

\subsection{Performance of Mitigated Application}
\label{sec:results-app}

Next, we evaluate the \Name{}-instrumented applications.  We focus on
the overhead in execution time and the change in compression ratio.
We present comparisons to other
approaches as scatter plots. Points above the 
dashed blue-line indicate \Name{} outperforms the 
competing approach.

\subsubsection{Execution Time}
We measure the overhead of our instrumented PHP code and modified
compressor separately, by comparing with the original (and leaky) PHP
code and original \textit{zlib} compressor~\cite{zlib}.  The results
are shown in Table~\ref{tbl:slowdown}.  Columns 3-5 compare the time in seconds
of both the original and instrumented PHP code.  Columns 6-8 compare
the time in milliseconds of the original and modified compressor code.  We average the PHP code
time over 100 executions, and the compressor code over 1000 executions
because it is a much faster process and therefore affected more by
noise.
%

For 22 of the 25 pages, the runtime overhead in PHP is $<$3\%, which
is negligible. This is because the time is generally dominated by
network or database accesses, which is also why the overhead is
occasionally negative, since the instrumentation overhead is small. 
In 24 out of 25 pages, the original and modified compressors are
within a $ 1/10^{th} $ of millisecond of each other, and for the
remaining page, \Name{} is faster. All compressor results are
statistically significant with $ p < 0.00001 $. \Name{} is faster when
the amount of sensitive data is large because Huffman coding is much
faster than searching for LZ77 matches. Finally, we compare \Name{}
with \emph{SafeDeflate} in Fig.~\ref{fig:sd_exec} and observe that \Name{} is
2-5 times faster.

\begin{table}

\caption{Runtime overhead of PHP code with compression.}

\label{tbl:slowdown}
\centering

\scalebox{0.765}{
\begin{tabular}{|l|c|c|c|r|c|c|r|}\hline
App & Page  & \multicolumn{3}{c|}{ PHP-script Running Time (s) } & \multicolumn{3}{c|}{ Compression Time (ms) } \\
 \cline{3-5}
 \cline{6-8}
   &  & \makecell{Orig.} & Instr. & Overhead  & \makecell{Orig.}  & Modif.  & Overhead \\\hline\hline
\multirow{5}{*}{Squirrelmail}
                              & comp-e & 0.00082 & 0.00082 & 0.00\% & 0.1115 & 0.1248 & 11.88\% \\
                              & login  & 0.00036 & 0.00036 & 0.00\% & 0.0428 & 0.0485 & 13.41\% \\
                              & pref.  & 0.00070 & 0.00068 & -2.86\% & 0.0830 & 0.0702 & -15.49\% \\
	                      		& view-e & 0.91540 & 0.98640  & 7.76\% & 0.2090 & 0.1594 & -23.74\% \\
                              &	view-i & 1.25098 & 1.25102 & 0.00\% & 0.2556 & 0.2275 & -11.00\% \\ \hline

\multirow{5}{*}{NOCC}         
                              & comp-e & 0.57718 & 0.55184 & -4.39\% & 0.0234 & 0.0258 & 10.62\% \\
                              & view-e & 3.45185 & 3.55438 & 2.97\% & 0.7342 & 0.6834 & -6.91\% \\
	                      & view-i & 6.61984 & 6.68215 & 0.94\% & 0.5270 & 0.5171 & -1.88\% \\
	                      & login  & 0.00067 & 0.00067 & 0.00\% & 0.1072 & 0.1172 & 9.34\% \\
	                      & pref.  & 0.59901 & 0.64006 & 6.85\% & 0.2428 & 0.2649 & 9.08\% \\ \hline

\multirow{5}{*}{Adminer}      
                              &	edit-r & 0.96962 & 0.97202 & 0.25\% & 0.1095 & 0.1231 & 12.41\% \\
	                      & isrt-i & 0.00212 & 0.00214 & 0.94\% & 0.0978 & 0.1143 & 16.87\% \\
	                      & login  & 0.00052 & 0.00053 & 1.92\% & 0.0303 & 0.0331 & 9.42\% \\
	                      & tbl-da & 0.46160 & 0.46117 & -0.09\% & 0.5788 & 0.6129 & 5.90\% \\
	                      & tbl-st & 0.00240 & 0.00247 & 2.92\% & 0.0893 & 0.1002 & 12.20\% \\ \hline

\multirow{5}{*}{iAddressbook} 
                              &	addr-b & 0.08181 & 0.08182 &  0.01\% & 0.2881 & 0.3128 & 8.58\% \\
	                      & c-info & 0.08213 & 0.08234 & 0.26\% & 0.3768 & 0.4248 & 12.72\% \\
	                      & edit-c & 0.08256 & 0.08244 &  -0.15\% & 0.6576 & 0.7498 & 14.01\% \\
	                      & login  & 0.00078 & 0.00076 &  -2.56\% & 0.0325 & 0.0346 & 6.49\% \\
	                      & new-c  & 0.08189 & 0.08195 &  0.07\% & 0.6465 & 0.7249 & 12.13\% \\ \hline

\multirow{5}{*}{Addressbook}  
                              &	c-info & 0.00195 & 0.00198 & 1.54\% & 0.1262 & 0.1423 & 12.79\% \\
	                      & edit-c & 0.00209 & 0.00213 & 1.91\% & 0.1982 & 0.2399 & 21.03\% \\
	                      & login  & 0.00092 & 0.00094 & 2.17\% & 0.0906 & 0.1009 & 11.38\% \\
	                      & new-c  & 0.00191 & 0.00195 & 2.09\% & 0.1144 & 0.1345 & 17.53\% \\
	                      & addr-b & 0.16089 & 0.17013 & 5.74\% & 7.8697 & 4.9878 & -36.62\% \\ \hline

\end{tabular}
}
\begin{tablenotes}
      \item \scriptsize $^*$  NOCC makes many connections to the email server and each connection latency varies every time, which explains the long execution time and big performance difference.
\end{tablenotes}
\end{table}

\subsubsection{Compression Ratio}

\begin{figure}
\centering
\scalebox{0.54}{\begin{tikzpicture}
\begin{axis}[
        xmin=0.1,
        ymin=0.1,
        xmax=0.7,
        ymax=0.7,
        ylabel=SafeDeflate,
        xlabel=Debreach
    ]
    \addplot[
        scatter,
        only marks,
        point meta=explicit symbolic,
        scatter/classes={
            sdvsdbr={ggreen},
            huffvsdbr={ppurple}
        }
    ] table [meta=label] {
x   y   label
0.262	0.453	sdvsdbr
0.444	0.528	sdvsdbr
0.289	0.419	sdvsdbr
0.426	0.531	sdvsdbr
0.427	0.531	sdvsdbr
0.250	0.358	sdvsdbr
0.258	0.371	sdvsdbr
0.444	0.527	sdvsdbr
0.463	0.401	sdvsdbr
0.267	0.390	sdvsdbr
0.201	0.426	sdvsdbr
0.200	0.390	sdvsdbr
0.140	0.302	sdvsdbr
0.375	0.425	sdvsdbr
0.136	0.298	sdvsdbr
0.467	0.639	sdvsdbr
0.579	0.511	sdvsdbr
0.235	0.552	sdvsdbr
0.330	0.567	sdvsdbr
0.220	0.551	sdvsdbr
0.345	0.488	sdvsdbr
0.432	0.596	sdvsdbr
0.347	0.476	sdvsdbr
0.554	0.557	sdvsdbr
0.409	0.474	sdvsdbr
    };
    \draw [bblue,dashed] (rel axis cs:0,0) -- (rel axis cs:1,1);
\end{axis}
\end{tikzpicture}%
\begin{tikzpicture}
\begin{axis}[
        xmin=0.1,
        ymin=0.1,
        xmax=0.7,
        ymax=0.7,
        ylabel=Huffman,
        xlabel=Debreach
    ]
    \addplot[
        scatter,
        only marks,
        point meta=explicit symbolic,
        scatter/classes={
            sdvsdbr={ggreen},
            huffvsdbr={ppurple}
        }
    ] table [meta=label] {
x   y   label
0.262	0.665	huffvsdbr
0.444	0.692	huffvsdbr
0.289	0.652	huffvsdbr
0.426	0.692	huffvsdbr
0.427	0.692	huffvsdbr
0.250	0.686	huffvsdbr
0.258	0.684	huffvsdbr
0.444	0.709	huffvsdbr
0.463	0.669	huffvsdbr
0.267	0.670	huffvsdbr
0.201	0.614	huffvsdbr
0.200	0.607	huffvsdbr
0.140	0.594	huffvsdbr
0.375	0.582	huffvsdbr
0.136	0.592	huffvsdbr
0.467	0.673	huffvsdbr
0.579	0.668	huffvsdbr
0.235	0.670	huffvsdbr
0.330	0.673	huffvsdbr
0.220	0.662	huffvsdbr
0.345	0.636	huffvsdbr
0.432	0.672	huffvsdbr
0.347	0.651	huffvsdbr
0.554	0.686	huffvsdbr
0.409	0.675	huffvsdbr

    };
    \draw [bblue,dashed] (rel axis cs:0,0) -- (rel axis cs:1,1);
\end{axis}
\end{tikzpicture}}
\caption{Comparison of compression ratios: \Name{} vs. \emph{SafeDeflate} (left) and \Name{} vs. Huffman-Only (right).\label{fig:comp_res}}
\end{figure}
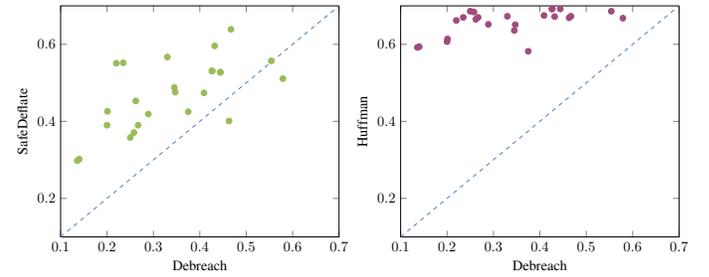

Fig.~\ref{fig:comp_res} compares the compression ratios
of \Name{}, \emph{SafeDeflate} and \emph{Huffman-Only}. 
Here, compression ratio is computed as $\frac{compressed \;
size}{original \; size}$.  
The $x$-axis shows the compression ratio for \Name{}, and the $y$-axis
shows the other. We note again here that \emph{SafeDeflate} does not even
provide security guarantees.

\Name{} is on average 15.6\% better than \emph{SafeDeflate}. This is 
because \emph{SafeDeflate} assumes any region of data composed of sensitive
alphabet characters is tainted, but non-sensitive data invariably
contain characters from the alphabet, so \emph{SafeDeflate}
over-approximates.
\Name{}, in contrast, is almost always more accurate.  For example, login and preference
pages either do not call sensitive API functions, or their results do
not flow into HTML response, which can be identified by \Name{}. Compared
to the oracle, \Name{} is on average within 2.5\%.

The reliance of \emph{SafeDeflate} on a sensitive alphabet is
especially problematic in NOCC, where sensitive data may contain
HTML. For example, on the inbox view page, \Name{} is able to
precisely taint only the email header information such as the
sender/receiver email addresses, the email subject lines, the sending
date, etc. This accounts for a relatively small portion of the data on
the inbox page. Unfortunately, \emph{SafeDeflate} determines full HTML
tags and URLs (which make up the majority of any HTML page) are
tainted as well. This leads to poor performance (more than 25\% worse
in four of five pages).



\subsection{Security Evaluation}
\label{sec:sec_eval}

Now, we report the exploitable leaks in two applications based on what
has been described in Section~\ref{sec:adv_model}.  That is, the
adversary is a man-in-the-middle and the victim visits an
adversary-controlled web page.  For each leak, we identify a target
secret to extract (for \textit{Squirrelmail}, it is one arbitrarily
chosen subject line from the inbox; for \textit{Adminer}, it is a
credit card number stored in the database).

\begin{figure}
\vspace{1ex}
\centering
\begin{minipage}{0.49\linewidth}
\centering
\includegraphics[width=\linewidth]{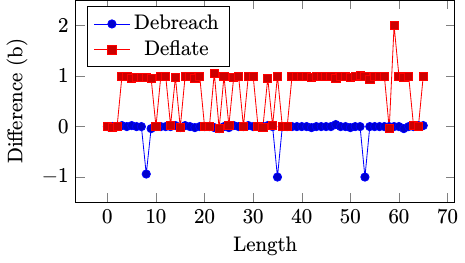}
{\scriptsize (a) Squirrelmail}
\end{minipage}
\begin{minipage}{0.49\linewidth}
\centering
\includegraphics[width=\linewidth]{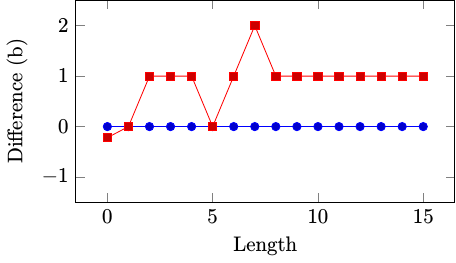}
{\scriptsize (b) Adminer}
\end{minipage}
\caption{Side-channel leak measurements in two case studies.\label{fig:leaks}}
\end{figure}

Fig.~\ref{fig:leaks} shows the ability to guess the $(n + 1)^{th}$ byte
of a target secret given that the attacker knows the first $n$ bytes.
The x-axis is the length $n$ of the known prefix, and the y-axis is
the difference between the compressed sizes with correct and incorrect
guesses of the $(n + 1)^{th}$ byte. Here, a difference near one or
greater indicates the leak is exploitable.

In both \textit{Squirrelmail} and \textit{Adminer}, there is a strong
potential for the application to be exploited once the adversary can
determine the first 3 bytes of the secret.  After \Name{} is applied,
however, the difference becomes less than or equal to zero, which
means the exploit is no longer possible.  In \emph{Squirrelmail}, the
correct guess is occasionally larger than incorrect guesses since the
encodings of Huffman coding changes, which explains the negative
difference. Therefore, this does not leak information.

\section{Related Work} 
\label{sec:related}

We have already reviewed techniques that are the most closely related
to our work, including various attacks that exploit the compression
side channel.

Some components of compression side channel attacks have
been leveraged in other types of attacks. For example, the timing
of cross-site requests have been exploited to reveal private
information in web applications \cite{van2015, gelernter2015,
bortz2007}. Chosen plaintext attacks have been previously used to
decrypt data encrypted using CBC-mode encryption
ciphers \cite{bard2006, joux2002, BEASTCVE}.
%
%
Recent work has been done on analyzing implementations of compression
routines \cite{detection1, detection2, synthesis}.
Zhou et al.~\cite{detection2}, in particular, use \textit{approximate}
model counting to estimate leakage, and are able to apply their
approach to a real-world compressor. However, none of these works can
mitigate the compression side channel in a server application, like we
do.

Besides compression, which is a relatively new source of exploitable
side channel information, other sources of side channel information
have been studied in the past such as timing\cite{antonopoulos2017,
chen2017, WuGS018, BrennanSBP18, WuW19}, cache behavior \cite{barthe2014, sung2018canal,
guo2018adversarial}, and power \cite{bayrak2013, EldibWS14tosem,
zhang2018, WangSW19}. However, attacks and mitigations for these side channels
differ fundamentally from compression side channels.

Datalog-based program analysis has been applied in many domains.  For
example, Whaley and Lam \cite{whaley2004} used this framework to
perform context-sensitive alias analysis in Java programs.  Livshits
and Lam~\cite{LivshitsL05} and Naik et al.~\cite{NaikAW06} used
similar techniques to detect security errors and data-races.
Bravenboer and Smaragdakis~\cite{BravenboerS09} used it to perform
points-to analysis.  
Sung et al.\ used it to improve automated testing
of JavaScript-based web applications~\cite{sung2016} and semantic
diffing of concurrent programs~\cite{SungLEW18}.
However, none of these prior works used Datalog to mitigate
compression side channels.

We also note here that we tried to use off-the-shelf solvers such
as \textit{Souffle}~\cite{souffle} to solve our Datalog
programs. While \textit{Souffle} finished in all cases, we found that
our Python-based solver finished much faster in some cases. This is
because our solver is geared toward iterative data-flow analysis (such
as our taint analysis) and only supports our rule set,
whereas \textit{Souffle} is general-purpose and supports arbitrary
rule sets.  An interesting research direction would be optimizing
general-purpose Datalog solvers such as \textit{Souffle} for iterative
data-flow analysis.

Finally, while not the focus of our work, there is a significant body
of work focusing on the static analysis of PHP programs. For example,
Xie and Aiken \cite{xie2006} are among the first to statically analyze
PHP to detect SQL injection vulnerabilities.  Since then, many have
proposed techniques to model features such as
aliasing~\cite{jovanovic2006}, built-in functions~\cite{dahse2014sim},
second-order data flows~\cite{dahse20142nd}, object
injection~\cite{dahse2014pop}, and client-side call
graphs~\cite{nguyen2014}. Most recently, Alhuzali et
al. \cite{alhuzali2018} combine static and dynamic analyses to
synthesize exploits such as SQL injection and cross-site scripting
attacks.

\section{Conclusions}
\label{sec:conclusion}

We have presented a \emph{safe} and \emph{efficient} compressor-level
approach to mitigating compression side channel attacks.  Our approach
is based on static taint analysis to safely find tainted sinks and
efficient code instrumentation techniques to instrument proper program
points.  It gives a server application the ability to automatically
generate annotations of sensitive data at run time.  Moreover, it is
fully compatible with existing platforms.  We have implemented our
approach in the software tool for PHP-based server applications and
showed that our approach is both efficient compared to
state-of-the-art mitigation techniques, and can prevent leaks on a set
of real-world applications, while having minor performance overhead.

\section*{Acknowledgments}

This work was partially funded by the U.S.\ National Science
Foundation (NSF) under the grant CNS-1617203 and Office of Naval
Research (ONR) under the grant N00014-17-1-2896.
%

\clearpage
\newpage
\bibliography{ref}

\end{document}